\documentclass[11pt]{article}
\usepackage{jcapmod}
\usepackage{rotating}
\usepackage{amsmath}
\usepackage{mathtools}
\usepackage{amsthm}
\usepackage{amsfonts}
\usepackage{graphicx}
\usepackage{psfrag}
\usepackage{hyperref}
\usepackage{amssymb}
\usepackage{lscape}
\usepackage{slashed}
\usepackage{lscape}
\usepackage{color}
\usepackage{url}
\usepackage{caption}
\usepackage{bm}
\usepackage{subfig}
\usepackage{epsfig}
\usepackage{soul}

\newcommand{\beq}{\begin{equation}}
\newcommand{\eeq}{\end{equation}}
\newcommand{\bea}{\begin{eqnarray}}
\newcommand{\eea}{\end{eqnarray}}
\newcommand{\ga}{\lower.7ex\hbox{$\;\stackrel{\textstyle>}{\sim}\;$}}
\newcommand{\la}{\lower.7ex\hbox{$\;\stackrel{\textstyle<}{\sim}\;$}}

\newcommand{\eff}{{\rm{ eff}}}
\newcommand{\bp}{\boldsymbol{p}}

\newcommand{\bx}{\boldsymbol{x}}

\newcommand*\diff{\mathop{}\!\mathrm{d}}
\newcommand{\CL}{\texttt{$\mathcal{C}\text{osmo}\mathcal{L}\text{attice}$}}
\newcommand{\overbar}[1]{\mkern 1.5mu\overline{\mkern-1.5mu#1\mkern-1.5mu}\mkern 1.5mu}

\newcommand{\Trh}{T_\text{RH}}

\newcommand{\arh}{a_\text{RH}}

\newcommand{\aend}{a_\text{end}}

\makeatletter
\newcommand{\Cen}[2]{%
  \ifmeasuring@
    #2%
  \else
    \makebox[\ifcase\expandafter #1\maxcolumn@widths\fi]{$\displaystyle#2$}%
  \fi
}
\makeatother

\begin{document}

\begin{flushright}
UMN--TH--4223/23, FTPI--MINN--23/15, DESY-23-122 \end{flushright}

\vspace{0.5cm}
\begin{center}
{\bf {\large Effects of Fragmentation on Post-Inflationary Reheating}}
\end{center}

\vspace{0.05in}

\begin{center}
{\bf Marcos~A.~G.~Garcia}$^{a}$,
{\bf Mathieu Gross}$^{b}$,
{\bf Yann Mambrini}$^{b}$,
{\bf Keith~A.~Olive}$^{c}$, 
{\bf Mathias Pierre}$^{d}$, and
{\bf Jong-Hyun Yoon}$^{b}$
\end{center}

\begin{center}
 {\em $^a$Departamento de F\'isica Te\'orica, Instituto de F\'isica, Universidad Nacional Aut\'onoma de M\'exico, Ciudad de M\'exico C.P. 04510, Mexico}\\[0.2cm] 
  {\em $^b$ Universit\'e Paris-Saclay, CNRS/IN2P3, IJCLab, 91405 Orsay, France}\\[0.2cm] 
 {\em $^c$William I. Fine Theoretical Physics Institute, School of
 Physics and Astronomy, University of Minnesota, Minneapolis, MN 55455,
 USA}\\[0.2cm] 
 {\em $^d$Deutsches Elektronen-Synchrotron DESY, Notkestr. 85, 22607 Hamburg, Germany}

\end{center}

\bigskip

\centerline{\bf ABSTRACT}

\noindent  
We consider the effects of fragmentation on the post-inflationary epoch of reheating. 
In simple single field models of inflation, an inflaton condensate undergoes an oscillatory phase once inflationary expansion ends. The equation of state of the condensate depends on the shape of the scalar potential, $V(\phi)$, about its minimum. Assuming $V(\phi) \sim \phi^k$, the equation of state parameter is given by $w = P_\phi/\rho_\phi = (k-2)/(k+2)$.
The evolution of condensate and the reheating process depend on $k$.  For $k \ge 4$, inflaton self-interactions may lead to the fragmentation of the condensate and alter the reheating process. Indeed, these self-interactions lead to the production of a massless gas of inflaton particles as $w$ relaxes to 1/3. If reheating occurs before fragmentation, the effects of fragmentation are harmless.  We find, however, that the effects of fragmentation depend sensitively to the specific reheating process. Reheating through the decays to fermions is largely excluded since perturbative couplings would imply that fragmentation occurs before reheating and in fact could prevent reheating from completion. Reheating through the decays to boson is relatively unaffected by fragmentation and reheating through scatterings results in a lower reheating temperature. 

\vspace{0.2in}

\begin{flushleft}
August 2023
\end{flushleft}
\medskip
\noindent

%\mg{MG} {\red YM} {\bf KO} {\green JY} {\color{cyan} MP}

\newpage

\hypersetup{pageanchor=true}

\tableofcontents

\newpage

%%%%%%%%%%%%%%%%%%%%%%%%%%%
\section{Introduction}
%%%%%%%%%%%%%%%%%%%%%%%%%%%

Subsequent to a sufficiently extended period of accelerated expansion in the early Universe 
known as inflation \cite{reviews}, a  `graceful exit' to a radiation dominated era must occur. This period of reheating was absent in 
early models of inflation based on a strongly first order phase transition \cite{Guth:1982pn}, but 
given a coupling of the inflaton to Standard Model fields, reheating in models of inflation based on the slow roll of the inflaton occurs quite naturally \cite{dl,nos}. 
In many simple models of inflation, the inflaton rolls to a minimum of the scalar potential which can be 
expanded as a quadratic potential about the minimum. In this case, the inflaton begins a period of 
damped harmonic oscillations during which the Universe expands as if it were matter dominated, i.e., with 
an equation of state parameter, $w = P_\phi/\rho_\phi = 0$. These oscillations continue until inflaton decay becomes possible, when the inflaton decay rate, $\Gamma_\phi$, exceeds the expansion rate, $H$, or more simply when the lifetime of the inflaton becomes less than the age of the Universe. 

This is of course an idealized picture of instantaneous reheating and thermalization. 
The production of radiation begins almost immediately after accelerated expansions ends and continues until $\Gamma_\phi \sim H$ \cite{Giudice:2000ex,Garcia:2017tuj,Chen:2017kvz,Kaneta:2019zgw,Bernal:2019mhf,gkmo1,Bernal:2020gzm,Co:2020xaf,gkmo2,mo,gsimp,Passaglia:2021upk,gkmov,cmov,Goudelis:2022bls,Mambrini:2022uol,Barman:2022qgt,Becker:2023tvd}. The radiation energy density quickly reaches a maximum and then redshifts more slowly than $a^{-4}$, where $a$ is the cosmological scale factor, as 
radiation is continuously being pumped into the bath as the inflaton decays.  Furthermore the
decay products are produced with a delta-function distribution and 
only through scatterings achieve a thermal distribution  \cite{Davidson:2000er,Harigaya:2013vwa,Mukaida:2015ria,GA,Passaglia:2021upk,Drees:2021lbm,Drees:2022vvn,Mukaida:2022bbo}. In this work we will be primarily interested in the 
evolution of the various components 
comprising the energy density and 
will assume instantaneous thermalization.

As already noted, in many models 
including the Starobinsky model of inflation \cite{staro}, the leading term in the expansion of the potential about the minimum is quadratic. 
Although the details of the reheating process are relatively insensitive to the specific inflationary model, they are sensitive to the shape of the inflaton potential around the minimum \cite{Bernal:2019mhf,gkmo1}. We assume that about the minimum, the potential can be expanded to give 
\beq\label{eq:Vphi}
V(\phi) \;=\; \lambda M_P^4 
\left(\frac{\phi}{M_P}\right)^k\,, \qquad \phi \ll M_P\,,
\eeq 
where the reduced Planck mass is $M_P=1/\sqrt{8\pi\,G}\simeq 2.4\times 10^{18}\,{\rm GeV}$.
As the universe expands during this period where the inflaton undergoes anharmonic oscillations (when $k > 2$), the equation of state parameter no longer behaves as matter, but rather $w = (k-2)/(k+2)$. For example, for $k=4$, $w = 1/3$ and the inflaton energy scales as radiation. More generally,
solving the Friedmann equation for $\rho_\phi$ gives \cite{Bernal:2019mhf,gkmo1,gkmo2}
\beq
\rho_\phi(a) = \rho_{\rm end} \left(\frac{a}{a_{\rm end}} \right)^{-\frac{6k}{k+2}} \, ,
\label{Eq:rhophin}
\eeq
where $\rho_{\rm end} = \rho_\phi(a_{\rm end})$ and $\aend$ corresponds to the value of the scale factor when accelerated expansion ends, defined when $\ddot a=0$, corresponding to $w = -1/3$ (recall that during inflation $w \simeq -1$).

Note that for $k > 4$, the inflaton energy density redshifts faster than radiation. It has been argued that in this case, fragmentation of the inflaton condensate becomes important and the equation of state rapidly evolves from some $w > 1/3$ to $w = 1/3$ and the universe is dominated by a gas of massless inflaton particles, redshifting as $a^{-4}$ \cite{Lozanov:2016hid,Lozanov:2017hjm}. These free particles are produced as a consequence of the self-interaction of the inflaton. Notably, even if $\lambda\ll 1$, as required by the measured amplitude of the primordial curvature fluctuation, the growth of inhomogeneities can be amplified by the accumulation of the (bosonic) field fluctuations over the oscillation of the inflaton. This parametric self-resonance persists despite the expansion of the universe, leading then to the exponential growth of non-zero momentum modes until they backreact into the homogeneous condensate, fragmenting it~\cite{Greene:1997fu,Kaiser:1997mp,Garcia-Bellido:2008ycs,Frolov:2010sz,Amin:2011hj,Hertzberg:2014iza,Figueroa:2016wxr,Lozanov:2016hid,Lozanov:2017hjm,Fu:2017ero,Antusch:2021aiw,Lebedev:2022vwf,Cosme:2022htl,Alam:2023kia,Garcia:2023eol}. The resulting inhomogeneous field distribution can become localized in transient structures, lasting for a few oscillations, or into long-lived, soliton-like structures called oscillons~\cite{Copeland:1995fq,Salmi:2012ta,Amin:2010dc,Amin:2011hj,Lozanov:2016hid,Lozanov:2017hjm}. If the fragmentation of the condensate is complete, the inflaton could no longer decay and barring sufficiently strong scattering to Standard Model particles, reheating becomes problematic. Here we show, that in fact, fragmentation while effective is never completed before inflaton decay. That is, some fraction of the inflaton energy density continues to reside in the condensate providing an effective inflaton mass and thus allowing for a decay. In this paper, we compute the non-perturbative effects leading to fragmentation and discuss the reheating process when these effects are taken into account. We derive the change in the final reheating temperature relative to the case when fragmentation effects are ignored. 

Our paper is organized as follows. In Section \ref{sec:gkmoreview}, we review the basic elements leading to reheating for arbitrary (even) $k$, assuming that reheating occurs from either the decay of the inflaton to fermions, bosons, for from scatterings to bosons \cite{gkmo2}. We also discuss the effects of kinematic blocking (or enhancement in the case of bosonic final states). In Section \ref{sec:analim}, we derive naive but analytical limits on reheating and fragmentation. We assume that fragmentation occurs instantly at a certain point in the evolution, and that it is total. That is when fragmentation occurs, there is a complete conversion of the energy density in the condensate into relativistic particles. In this case, reheating will not occur if the decay rate is too small and decays have not completed before fragmentation. Thus we can set a lower limit on the inflaton decay/scattering coupling in terms of epoch of fragmentation. Conversely, for a given coupling we set a limit on the duration of oscillations before fragmentation begins. 
In Section \ref{sec:frag}, we perform numerical simulations showing the onset of fragmentation and the evolution of the condensate and the energy density of inflaton radiation. 
We see that while fragmentation is nearly complete, roughly a few percent of the condensate remains, and the interaction between the free particles and the condensate is enough to allow decay to occur and reheating to complete. We then revise the resulting reheat temperatures derived when fragmentation effects are included. 
Our conclusions are summarized in Section \ref{sec:disc}.

%%%%%%%%%%%%%%%%%%%%%%%%%%%
\section{Reheating through decay and scattering}
\label{sec:gkmoreview}
%%%%%%%%%%%%%%%%%%%%%%%%%%%

Perturbative reheating can often be treated analytically. 
We review in this section the analytical estimations for the reheating temperature from inflaton decay and scattering, following the analysis of \cite{gkmo2}. We consider potentials of the form given in Eq.~(\ref{eq:Vphi}). Of course,
single polynomial potentials are no longer viable as inflationary models due to CMB precision measurements of the tilt of the anisotropy spectrum, $n_s$ and the ratio of tensor to scalar perturbations, $r$ \cite{Planck}. However, about the origin, such potentials may arise from  $\alpha$-attractor T-models~\cite{Kallosh:2013hoa} 
\beq\label{eq:attractor}
V(\phi) \;=\; \lambda M_P^4 \left[ \sqrt{6} \tanh \left(\frac{\phi}{\sqrt{6}M_P}\right) \right]^k\, ,
\eeq
which can be derived simply in the context of no-scale supergravity~\cite{gkmo1}. In our analysis we normalized the potential with
\begin{equation}
    \lambda \, = \, \frac{18 \pi ^2 A_s}{6^{k/2} N_*^2}
    \label{eq:normalizationlambda}
\end{equation}
with $A_s=e^{3.044}/10^{10}$ being the amplitude of the scalar power spectrum and we chose $N_*=55$ as the number of e-folds between the CMB fiducial scale $k_*=0.05~\text{Mpc}^{-1}$ crossing and the end of inflation, in good agreement with current data \footnote{A more accurate treatment of the normalization and its relation to the number of e-foldings and reheat temperature for these models was discussed in \cite{egnov}}.  Inflation occurs at large ($\phi > M_P$) field values, and the end of inflation may be defined when $\ddot a=0$ where $a$ is the cosmological scale factor. Subsequently, the inflaton condensate begins a period of oscillations. For the potential in Eq.~(\ref{eq:attractor}), the inflaton field value marking the end of inflation is given by \cite{Ellis:2015pla, gkmo2}
\begin{align}
   \phi_{\rm end} &\simeq 
   \sqrt{\frac{3}{8}}M_P\ln\left[\frac{1}{2} + \frac{k}{3}\left(k + \sqrt{k^2+3}\right)\right].
   \label{phiend}
\end{align}
Inflaton oscillations continue until either decays or scatterings produce enough radiation (e.g. relativistic decay products) so that energy density in the newly formed radiation bath comes to dominate the total energy density and hence the expansion rate of the Universe. We define reheating to occur at the moment when $\rho_\phi  = \rho_R$,
where $\rho_\phi$ is the energy density stored in the inflaton oscillations, and $\rho_R$ as the energy density in radiation (assumed here to thermalize instantaneously \footnote{See \cite{GA} for the consequences of dropping this assumption.}).

\subsection{Reheating}
 
The case of $k=2$, is perhaps the best studied of this class of models \footnote{The Starobinsky 
model \cite{staro} also leads to a 
potential of the form in Eq.~(\ref{eq:Vphi}) with $k=2$}. In this case, the inflaton has a constant mass with $m^2_\phi = 2 \lambda M_P^2$, and the energy density of the inflaton
scales as $a^{-3}$, just as matter with an equation of state parameter $w = 0$. The premature destruction or fragmentation of the condensate would lead 
to massive inflaton quanta at rest which decay (or scatter) to reheat the 
universe leaving the qualitative picture intact. That is, the equation of state remains at $w=0$.
Similarly for $k=4$, 
the energy density in oscillations scale as $a^{-4}$, 
typical of radiation with $w=1/3$. In this case the 
destruction of the condensate leads to a bath of massless 
inflaton quanta with an unchanged equation of state \cite{Garcia:2023eol}. 
However if the condensate is completely destroyed, decays 
are no longer possible due to the absence of a mass for 
the $\phi$-particles, and reheating relies on the possibility of inflaton scattering. We will return to this point below.

For $k>4$, the effects of fragmentation can be more pronounced as the equation of state changes from $w > 1/3 \to w=1/3$ \cite{Lozanov:2016hid,Lozanov:2017hjm}. The equation of state parameter is determined from the inflaton equation of motion
\beq\label{eq:phieom}
\ddot{\phi} + (3H+\Gamma_{\phi})\dot{\phi} + V'(\phi) \;=\;0 \, ,
\eeq
where we have included the effects of inflaton decay with rate, $\Gamma_\phi$, and $H=\frac{\dot a}{a}$ is the Hubble parameter.
This can be rewritten in terms of the energy density and pressure stored in the scalar field
\beq
\rho_\phi = \frac{1}{2} \dot \phi^2 + V(\phi);
~~~P_{\phi} = \frac{1}{2} \dot \phi^2 - V(\phi) \, ,
\label{Eq:rhophi0}
\eeq
giving \cite{gkmo2}
\beq
\dot{\rho}_{\phi} + 3H(1+w)\rho_{\phi} \;\simeq\; -\Gamma_{\phi}(1+w)\rho_{\phi}\,.
\label{Eq:conservation}
\eeq
In writing Eq.~(\ref{Eq:conservation}),
we have assumed that after inflation, $\phi(t) = \phi_0(t)\cdot \mathcal{P}(t)$,
where the function $\mathcal{P}(t)$ is quasi-periodic and encodes the (an)harmonicity of the short time-scale oscillations in the potential and
the envelope $\phi_0(t)$ encodes the effect of redshift and decay, and varies on longer time-scales. If the oscillation time-scale is much shorter than the decay and redshift time-scales, we can average the energy density and pressure over an oscillation and find \cite{gkmo2}
\beq\label{eq:wk}
w \;=\; \frac{k-2}{k+2}\,,
\eeq
and the mean energy density averaged over the oscillations is
\beq
\langle\rho_\phi\rangle\;=\;V(\phi_0)\,.
\label{Eq:meanrhophi}
\eeq
The  radiation density produced by inflaton decay or scattering
is
\beq
\dot{\rho}_{R} + 4H \rho_{R} \;\simeq\; (1+w)\Gamma_{\phi}\rho_{\phi}\, , 
\label{eq:aeom}
\eeq
which together with the Friedmann equation
\beq
\rho_{\phi}+\rho_{R} \;=\; 3H^2 M_P^2\,,
\label{hub}
\eeq 
allows one to solve for $\rho_\phi(t), \rho_R(t)$, and $a(t)$ simultaneously and effectively for $\rho_\phi(a)$
and $\rho_R(a)$. 
The reheating temperature is then determined from 

\beq
\frac{g_\rho \pi^2}{30} \Trh^4 = \rho_R(\arh) \, ,
\eeq
when $\rho_\phi(\arh) = \rho_R(\arh)$. Here $g_{\rho}$ is the number of relativistic thermal degrees of freedom at reheating.

\begin{figure}[!ht]
\centering
    \includegraphics[width=\textwidth]{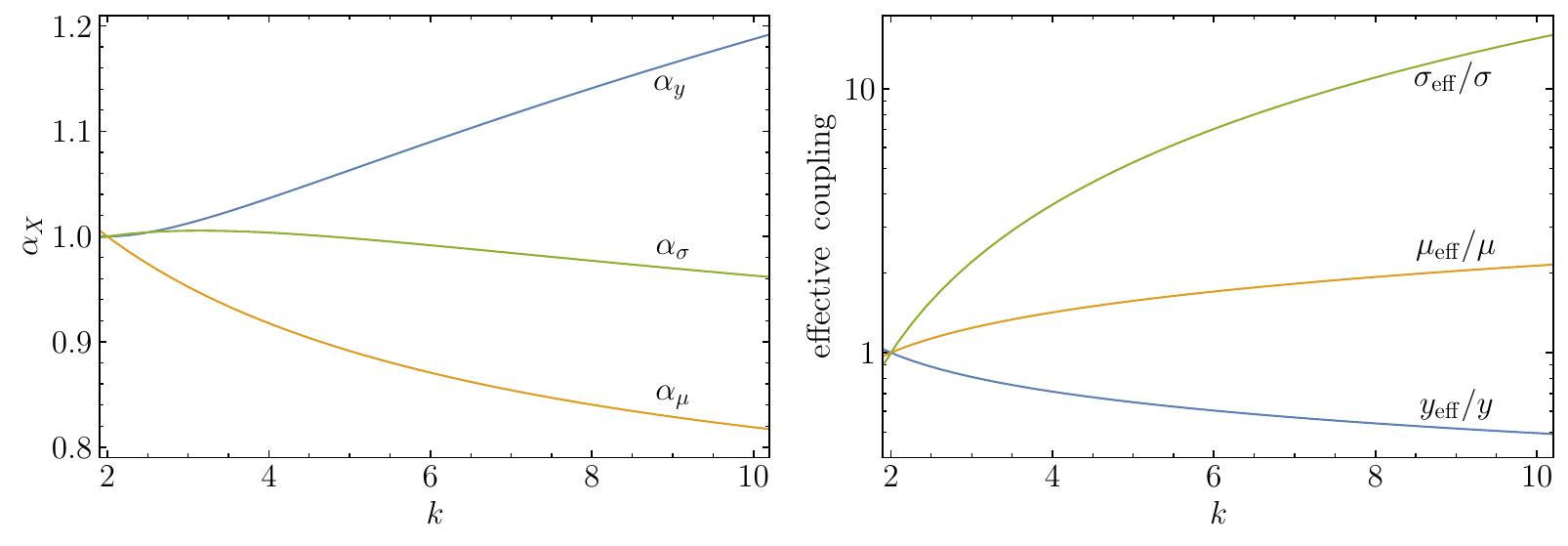}
    \caption{Left: The coefficient determining the effective couplings when kinematic blocking/enhancement effects are ignored, $\alpha_X(k,0)$, for $X = y, \mu, \sigma$. For $\mathcal{R} > 1$, $\alpha_{y,\sigma}(k,\mathcal{R} )\sim \alpha_{y,\sigma}(k,0) \mathcal{R}^{-1/2}$ and $\alpha_{\mu}(k,\mathcal{R} )\sim \alpha_{\mu}(k,0) \mathcal{R}^{1/2}$.  Right: The effective couplings relative to the Lagrangian couplings as a function of $k$ when $\mathcal{R} = 0$. 
    }
    \label{alphaplot}
\end{figure}

As discussed in detail in \cite{gkmo2}, 
the dynamics of reheating depend not only on $k$,
but also on whether 
the inflaton decays to fermions, bosons or scatters into bosons.
As the limits from fragmentation we will derive in the next section will also depend on the reheating modes, we will consider that the Lagrangian contains one of the three following couplings of the inflaton to matter
\beq
\mathcal{L}\supset  \begin{cases} 
y \phi \bar{f}f & \phi \to \bar{f}f \\
\mu \phi bb & \phi \to b b \\
\sigma \phi^2b^2 &  \phi \phi \to b  b ,
\end{cases}
\label{couplings}
\eeq
with $f\,(b$) standing for a fermionic (bosonic) final state. The decay (scattering) rate for each these modes is easily calculated
\beq
\Gamma_\phi = \begin{cases} 
 \frac{y_{\eff}^2(k)}{8\pi}  m_{\phi}(t)  & \phi \to \bar{f}f \\
\frac{\mu_{\eff}^2(k)}{8\pi  m_{\phi}(t) } & \phi \to b b \\
\frac{\sigma_{\eff}^2}{8 \pi} 
\frac{\rho_\phi(t)}{ m^3_\phi(t)} &  \phi \phi \to b  b .
\end{cases}
\label{couplings2}
\eeq
The effective couplings $y_{\eff}$, $\mu_{\eff}$ and 
$\sigma_{\eff}$ appearing in Eq.~(\ref{couplings2}) are each proportional to the couplings in Eq.~(\ref{couplings}) with the constant of proportionality depending on $k$ and will be discussed further in the next subsection. We show their dependence on $k$ in the right panel of Fig.~\ref{alphaplot}.
Their {\it effective} nature arises from the fact that the production of radiation depends on the details of the oscillations. Only for $k=2$, is the naive notion of a particle decay appropriate in the current context. For more details, see \cite{gkmo2}.  

Noting that the inflaton mass, $m_\phi(t) \propto \phi_0(t)^{(k-2)/k}$,
we can rewrite the inflaton mass as $m_\phi=\sqrt{k(k-1)}\lambda^{1/k}\rho^{(k-2)/2k}$. Then, averaging over several oscillations
for each of these modes 
to obtain
\beq
\Gamma_{\phi} \;=\;
\gamma_{\phi}\left(\frac{\rho_{\phi}}{M_P^4}\right)^{l}\,,
\label{Eq:gammaphi}
\eeq
where
\beq
\label{Eq:gammaphibis}
\gamma_{\phi} \;=\; 
\begin{cases}
\sqrt{k(k-1)}\lambda^{1/k}M_P\dfrac{y_{\eff }^2}{8\pi}\,,\quad & \phi\rightarrow \bar{f}f\,,\\[10pt]
\dfrac{\mu_{\eff }^2}{8\pi\sqrt{k(k-1)}\lambda^{1/k}M_P}\,,\quad& \phi\rightarrow bb\,,\\[10pt]
\dfrac{\sigma_{\eff}^2 M_P}{8\pi [k(k-1)]^{3/2}\lambda^{3/k}}\,,\quad & \phi\phi\rightarrow bb\,,
\end{cases}
\eeq
and
\beq
l \;=\; 
\begin{cases}
\frac{1}{2}-\frac{1}{k}\,,\quad & \phi\rightarrow \bar{f}f\,,\\
\frac{1}{k}-\frac{1}{2}\,,\quad & \phi\rightarrow bb\,,\\
\frac{3}{k}-\frac{1}{2}\,,\quad & \phi\phi\rightarrow bb\, ,
\end{cases}
\label{ls}
\eeq
and the value of $l$ simply reflects the power of $m_\phi$ in the various decay (scattering) channels. 

Using Eq.~(\ref{Eq:gammaphibis}), the solution to Eq.~(\ref{Eq:conservation}) is simply
\beq
\rho_\phi(a) = \rho_{\rm end} \left(\frac{a}{a_{\rm end}} \right)^{-\frac{6k}{k+2}} \, ,
\label{Eq:rhophi}
\eeq
valid when $\Gamma_\phi \ll H$. Using this solution, we can integrate Eq.~(\ref{eq:aeom})
giving
\beq
\rho_R(a) \; \simeq \; \frac{2 k}{(k + 8 - 6kl)} \frac{\gamma_\phi}{H_{\rm end}}
 \frac{\rho_{\rm end}^{l+1}}{M_P^{4l}} \left( \frac{a_{\rm end}}{a} \right)^{\frac{3k + 6kl}{k+2}} \, ,
\label{Eq:rhoRapp}
\eeq
valid when
 $8+k-6kl>0$ and
$a\gg a_{\rm end}$.
These two expressions determine the `moment' of reheating $\rho_\phi(\arh)=\rho_R(\arh)$ as 
\beq
\frac{a_{\rm RH}}{a_{\rm end}}=\left[ \frac{k+8-6kl}{2k} \frac{M_P^{4l-1}\rho_{\rm end}^{\frac{1}{2}-l}}{\sqrt{3}\gamma_\phi}\right]^{\frac{k+2}{3k-6kl}} \, ,
\eeq
and 
\begin{align}
T_{\rm RH} \;&=\; \left(\frac{30}{g_\rho \pi^2} \right)^{\frac{1}{4}}
\left[\frac{2k}{k+8-6kl} 
\frac{\sqrt{3}\gamma_\phi}{M_P^{4l-1}} \right]^{\frac{1}{2-4l}} \, .
\label{trh1}
\end{align}

The analogous expressions when $8+k-6kl<0$ are
\beq
\rho_R^{a\gg a_{\rm end}} \;=\; \frac{2k}{(6kl-k-8)} \frac{\gamma_\phi}{H_{\rm end}} \frac{\rho_{\rm end}^{l+1}}{M_P^{4l}}  \left(\frac{a_{\rm end}}{a}\right)^{4}\,,
\label{Eq:rhoRappter}
\eeq
so that $T \sim a^{-1}$. We then obtain

\beq
\frac{a_{\rm RH}}{a_{\rm end}}=\left[ \frac{6kl-k-8}{2k} \frac{M_P^{4l-1}\rho_{\rm end}^{\frac{1}{2}-l}}{\sqrt{3} \gamma_\phi}\right]^{\frac{k+2}{2k-8}} \, .
\label{arhaendneg}
\eeq
and (for $k>4$)
\begin{align}
T_{\rm RH} \;&=\; \left(\frac{30}{g_\rho \pi^2} \right)^{\frac{1}{4}}
\left[\frac{2k}{6kl-k-8} 
\frac{\sqrt{3}\gamma_\phi}{M_P^{4l-1}} \rho_{\rm end}^{\frac{6kl-k-8}{6k}}\right]^{\frac{3k}{4k-16}} \, .
\label{trh1-}
\end{align}

\subsection{Kinematic blocking}

\begin{figure}[!ht]
\centering
    \includegraphics[width=0.95\textwidth]{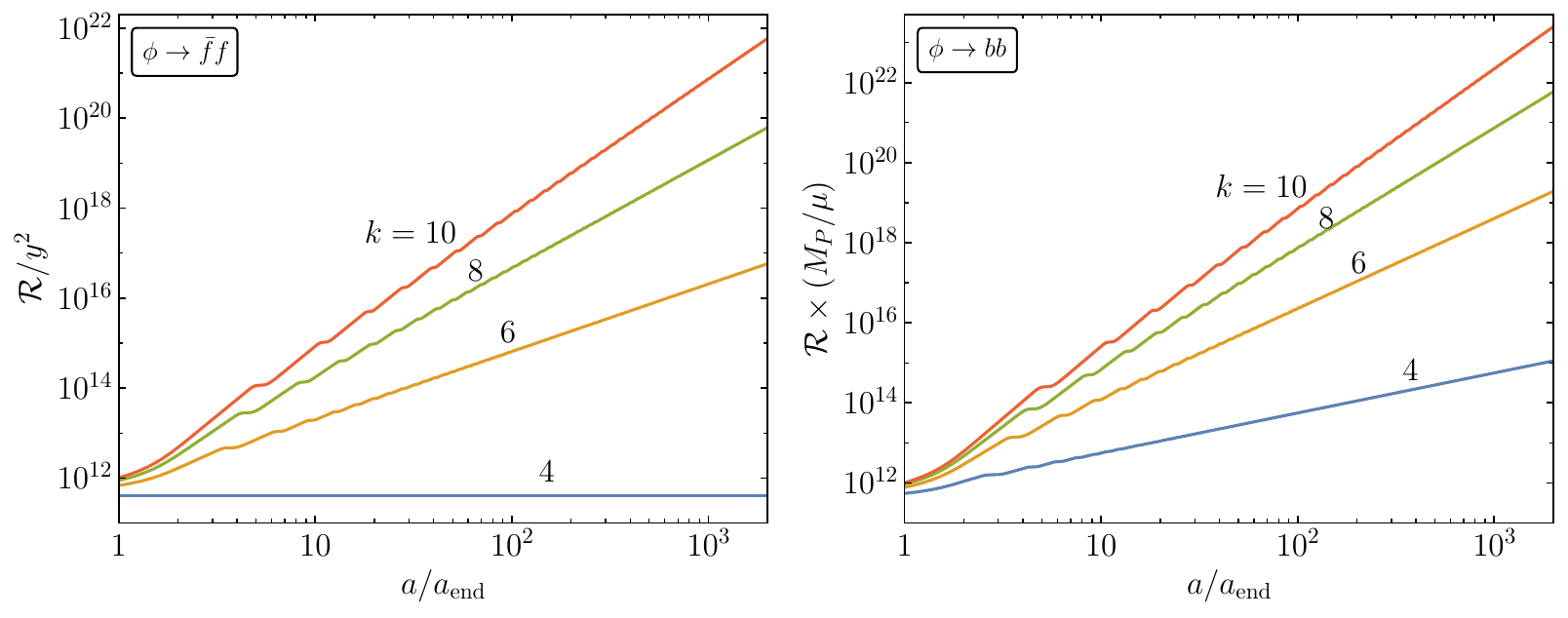}
    \caption{Kinematic parameter $\mathcal{R}$ as a function of the scale factor, for $k=4,6,8,10$. Left: fermionic decays. The channel $\phi\phi\rightarrow bb$ can be recovered from these results upon changing $y^2\rightarrow 2\sigma$. Right: bosonic decays.}
    \label{fig:R}
\end{figure}

The particle production rates evaluated above are in part determined by the time-dependent effective mass of the inflaton during the coherent oscillation of its background value $\phi(t)$. However, the form of the inflaton-matter couplings (\ref{couplings}) implies that also the latter acquires an induced mass. Namely,
\beq
m_{\rm eff}^2(t) \;\equiv\; \begin{cases}
y^2\phi^2\,, & \phi\rightarrow\bar{f}f\,,\\
2\mu\phi\,, & \phi\rightarrow bb\,,\\
2\sigma\phi^2\,, & \phi\phi\rightarrow bb\,.
\end{cases}
\eeq
As a consequence, in perturbation theory, the condition $m_{\rm eff}(t) \ll m_{\phi}$ is necessary for an efficient inflaton decay. Careful evaluation of the $\phi$ decay rates by means of the Boltzmann approximation shows that the parameter which determines the strength of this kinematic effect is~\cite{gkmo2}
\beq
\mathcal{R} \;\equiv\; \frac{8}{\pi k^2\lambda}\left(\frac{\Gamma(\frac{1}{k})}{\Gamma(\frac{1}{2}+\frac{1}{k})}\right)^2\times \begin{cases}
y^2\left(\dfrac{\phi_0(t)}{M_P}\right)^{4-k}\,, & \phi\rightarrow\bar{f}f\,,\\
2\dfrac{\mu}{M_P}\left(\dfrac{\phi_0(t)}{M_P}\right)^{3-k}\,, & \phi\rightarrow bb\,,\\
2\sigma \left(\dfrac{\phi_0(t)}{M_P}\right)^{4-k}\,, & \phi\phi\rightarrow bb\,,
\end{cases}
\eeq
which is $\mathcal{R}\propto(m_{\rm eff}/m_{\phi})^2_{\phi\rightarrow\phi_0}$. In the case of fermionic decays, or scattering depletion, the decay rate acquires a correction $\Gamma_{\phi}\propto \mathcal{R}^{-1/2}$ if $\mathcal{R}\gg 1$, resulting in a reduced efficiency of the inflaton decay. On the other hand, for $\phi\rightarrow bb$, due to the tachyonic nature of the effective mass of $b$ during half of the inflaton oscillation, an enhancement of the dissipation rate appears, $\Gamma_{\phi}\propto \mathcal{R}^{1/2}$ for $\mathcal{R}\gg 1$.
Fig.~\ref{fig:R} shows the scale factor dependence during reheating of the kinematic parameter. As is clear, only for very small couplings can this effect be neglected. For example, for fermionic decays $\mathcal{R}<1$ at the beginning of reheating only if $y\lesssim 10^{-6}$. Such a Yukawa coupling ensures that kinematic blocking can be neglected throughout reheating for $k=4$, but for higher $k$ the $\mathcal{R}>1$ regime is always reached because in this case the inflaton mass redshifts faster than the fermion mass.

The effective couplings used in Eq.~(\ref{Eq:gammaphibis}) can be defined in terms of the kinematic blocking/enhancement. 
For fermion final states
\beq\label{eq:yeffdef}
y_{\rm eff}^2(k) \;=\; \alpha_y(k,\mathcal{R})\frac{\omega}{m_{\phi}} y^2\, ,
\eeq
where $\alpha_y$ is determined by averaging the decays over many oscillations and $\omega$ is the frequency of oscillations \cite{gkmo2},
\beq
\frac{\omega}{m_\phi}= \sqrt{\frac{\pi k}{2(k-1)}}\,\frac{\Gamma(\frac{1}{2}+\frac{1}{k})}{\Gamma(\frac{1}{k})}\,.
\eeq
Similarly, 
\beq 
\mu_{\rm eff}^2(k) 
\equiv\; \frac{1}{4}(k+2)(k-1)\frac{\omega}{m_{\phi}} \alpha_{\mu}(k,\mathcal{R}) \mu^2\, ,
\label{eq:zeffdef}
\eeq 
and
\beq
\sigma^2_{\rm eff}(k) 
\equiv\; \frac{1}{8} k(k+2)(k-1)^2 \frac{\omega}{m_{\phi}} \alpha_{\sigma}(k,\mathcal{R}) \sigma^2\,.
\label{eq:seffdef}
\eeq
The effective couplings are shown in the right panels of Fig.~\ref{alphaplot}
and the left panels show the behavior of 
$\alpha_X(k,0)$, i.e., when the kinematic blocking factors are neglected. As noted earlier when included, $y_{\rm eff}^2$ and $\sigma_{\rm eff}^2$ each scale as $\mathcal{R}^{-1/2}$, whereas $\mu_{\rm eff}^2$ scales as $\mathcal{R}^{1/2}$.

\begin{figure}[!ht]
\centering
\includegraphics[width=0.90\textwidth]{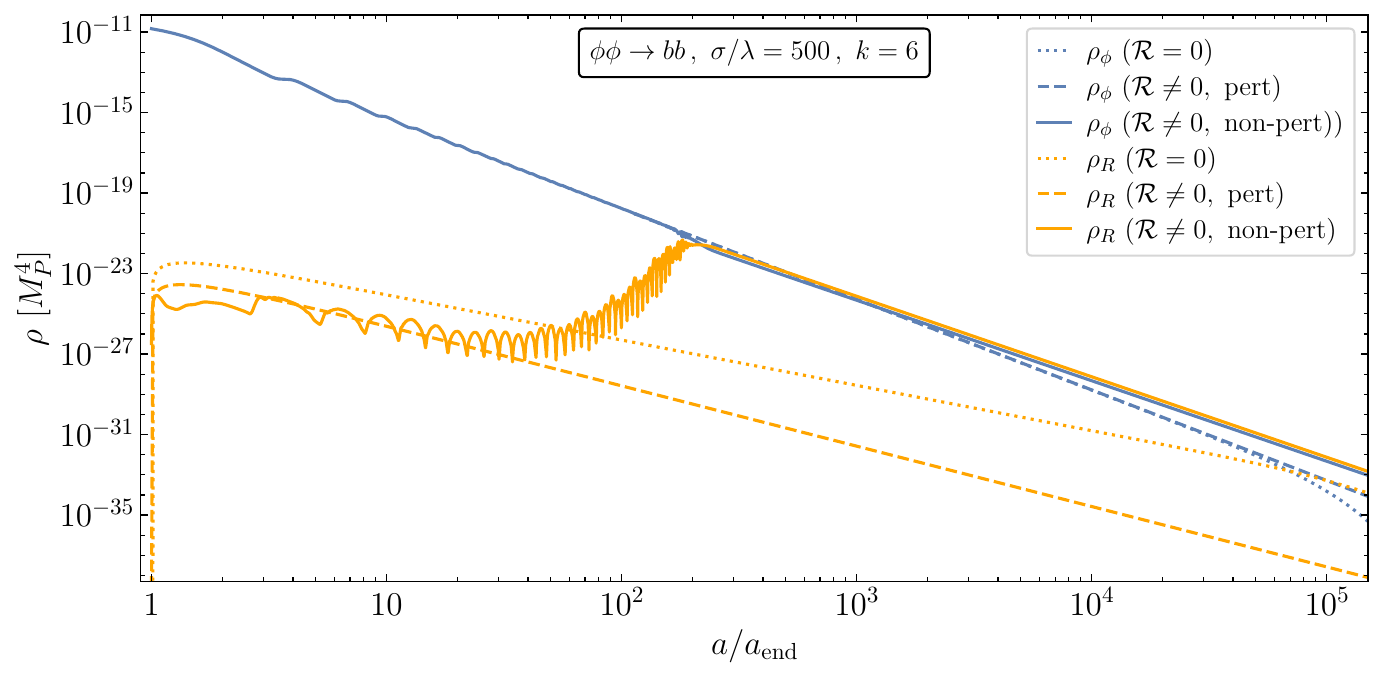}
    \caption{Dependence on the induced mass $m_{\rm eff}$ of the inflaton and radiation energy densities, for the $\phi\phi\rightarrow bb$ decay channel. The solid lines are computed by matching results using the Hartree approximation prior to strong parametric resonance to results from a lattice simulation for the backreaction regime.} %
    \label{fig:Rrho}
\end{figure}

Beyond perturbation theory, $\mathcal{R}\gg 1$ signals the presence of bosonic enhancement/fermionic blocking due to the appearance of strong parametric resonance. In particular, in the case of scalar $b$ production, the resonant growth will accumulate over the inflaton oscillation due to the non-adiabatic change of the effective mass, leading to exponential particle production. Qualitatively, the result is a transient kinematic blocking followed by a non-perturbative growth of the scalar energy density, as shown in Fig.~\ref{fig:Rrho}. Such non-perturbative excitations of decay products could in principle also contribute to the fragmentation of the inflaton. This is expected to be important in the case of decays to relatively strongly coupled scalars, prone to parametric resonance. Given the fact that the resonance structure strongly depends on the coupling between the inflaton and its decay products, one would need to perform a full analysis for each possible value of this coupling. In our analysis, we focus on post-fragmentation reheating and in order to remain as general as possible, such effects are disregarded here and left for future analyses.

\section{Fragmentation: an analytical approach}
\label{sec:analim}

We are now in a position to estimate the effects of fragmentation on the picture 
outlined in the previous section. For now, we will simply assume that the 
condensate is completely destroyed when the scale factor takes the value $a_\beta$ and we define $\beta \equiv a_\beta/\aend$. $\beta$ will be determined in the next section numerically for some specific examples. 
Naively, in the case of inflaton decay to either fermions or bosons, for $k \ge 4$, we must require $\beta > \arh/\aend$. Otherwise,
the destruction of the condensate will leave the universe with a bath of dark radiation (made of stable massless inflatons), and Standard Model radiation both with energy densities which scale as $a^{-4}$
and with $\rho_\phi > \rho_R$
which is clearly unacceptable. However, in the case of scattering, even if the particles are massless, scatterings may still occur and reheating may still be possible albeit with an altered reheating temperature. Note that the effects of fermionic kinematic blocking or bosonic enhancement are taken into account here by using the effective couplings $y_{\rm eff}$, $\mu_{\rm eff}$
and $\sigma_{\rm eff}$.

\subsection{Decay to fermions}
\label{Sec:decaytofermions}

Let us first consider the case of decay to fermions.
The condition that fragmentation occurs after reheating for  $8+k-6kl>0$ becomes
\beq
\beta > \left[ \frac{8 \pi (7-k)}{\sqrt{3} \lambda^\frac1k k \sqrt{k(k-1)} y_{\rm eff}^2} \left(\frac{\rho_{\rm end}}{M_P^4} \right)^\frac1k \right]^{\frac{k+2}{6}} \, ,
\label{blim}
\eeq
or
\beq
y_{\rm eff} > \left[ \frac{8 \pi (7-k)}{\sqrt{3} \lambda^\frac1k k \sqrt{k(k-1)}} \left(\frac{\rho_{\rm end}}{M_P^4} \right)^\frac1k \right]^{\frac12} \beta^{-\frac{3}{k+2}}
\label{ylim}
\eeq
and for  $8+k-6kl<0$
\beq
\beta > \left[ \frac{8 \pi (k-7)}{\sqrt{3} \lambda^\frac1k k \sqrt{k(k-1)} y_{\rm eff}^2} \left(\frac{\rho_{\rm end}}{M_P^4} \right)^\frac1k \right]^{\frac{k+2}{2k-8}} \, ,
\label{blim-}
\eeq
or
\beq
y_{\rm eff} > \left[ \frac{8 \pi (k-7)}{\sqrt{3} \lambda^\frac1k k \sqrt{k(k-1)}} \left(\frac{\rho_{\rm end}}{M_P^4} \right)^\frac1k \right]^{\frac12} \beta^{-\frac{k-4}{k+2}}
\label{ylim-}
\eeq
For the specific example of $k=6$, we have \cite{gkmo2,cmov}
$\lambda = 5.7 \times 10^{-13}$ and $\rho_{\rm end}^{1/4} = 4.9 \times 10^{15}$ GeV, implying that
\beq
\beta > 0.71\, y_{\rm eff}^{-8/3} \qquad {\rm or} \qquad 
y_{\rm eff} > 0.88\,\beta^{-3/8}
\label{bylimf}
\eeq
We can combine either Eqs.~(\ref{trh1}) and (\ref{ylim}) or Eqs.~(\ref{trh1-}) and (\ref{ylim-}) to obtain a limit on the reheating temperature which is also obtained from $\rho_{\rm RH}=\rho_R(\arh)>\rho_\phi(a_\beta)=\rho_{\rm end}\left(\frac{a_{\rm end}}{a_\beta}\right)^{\frac{6k}{k+2}}$
\beq
\Trh > \left(\frac{30}{g_\rho \pi^2} \right)^{\frac{1}{4}} \left(\frac{\rho_{\rm end}}{M_P^4} \right)^\frac14 \beta^{-\frac{3k}{2(k+2)}} M_P \, ,
\label{Eq:minimaltrhfermions}
\eeq
which for $g_\rho = 100$ and $k = 6$
is
\beq
\Trh > 2.0 \times 10^{15} \beta^{-\frac{9}{8}}~\rm{GeV} \simeq 6.5 \times 10^{10} \left(\frac{10^4}{\beta}\right)^{\frac{9}{8}}\rm{GeV}\, .
\label{trhbe}
\eeq
So long (\ref{trhbe}) is satisfied, there is no harmful effect of fragmentation.
Note however 
that our analytical result 
has been obtained in the very conservative limit that the inflaton condensate disappeared instantaneously
under the effect of fragmentation, 
which is not exactly the case as we will
see in a more precise numerical 
analysis. If residues remain after fragmentation, they will allow 
condensate to continue the reheating process, however, lowering the reheating temperature.

\subsection{Decay to scalars}

In this case, the constraint $\beta > \arh/\aend$ gives
\beq
\beta > \left[ \frac{8 \pi (2k+1)\sqrt{k(k-1)}\lambda^\frac1k M_P^2}{\sqrt{3} k  \mu_{\rm eff}^2} \left(\frac{\rho_{\rm end}}{M_P^4} \right)^{(1-\frac1k)} \right]^{\frac{k+2}{6k-6}} \, ,
\label{blims}
\eeq
or
\beq
\mu_{\rm eff} > \left[ \frac{8 \pi (2k+1)\sqrt{k(k-1)}\lambda^\frac1k M_P^2}{\sqrt{3} k  } \left(\frac{\rho_{\rm end}}{M_P^4} \right)^{(1-\frac1k)} \right]^\frac12 \beta^{-\frac{3k-3}{k+2}} \, ,
\label{mlims}
\eeq
which for $k=6$ reduces to
\beq
\beta > 2.9 \times 10^7  (\mu_{\rm eff}/{\rm GeV})^{-8/15} \qquad {\rm or} \qquad \mu_{\rm eff} > 9.8\times 10^{13} \beta^{-15/8} \, {\rm GeV}\, ,
\eeq
where $\mu_{\rm eff}$ is given in GeV.
The limit on the reheating temperature is again given by
the condition $\rho_{\rm RH}>\rho_{\phi}(a_\beta)$,
Eq.~(\ref{trhbe}).

\subsection{Scatterings to bosons}

The final case we consider is the scattering of inflatons to two bosons.
While we can still derive a limit on $\beta$ such that fragmentation does not affect the reheating process, reheating may still proceed via scattering if the limit is violated.
For fragmentation to occur after reheating, we need
\beq
\beta > \left[ \frac{8 \pi (2k-5)(k(k-1))^\frac32 \lambda^\frac3k }{\sqrt{3} k  \sigma_{\rm eff}^2} \left(\frac{\rho_{\rm end}}{M_P^4} \right)^{(1-\frac3k)} \right]^{\frac{k+2}{6k-18}} \, ,
\label{blimsc}
\eeq
or
\beq
\sigma_{\rm eff} > \left[ \frac{8 \pi (2k-5)(k(k-1))^\frac32 \lambda^\frac3k }{\sqrt{3} k  } \left(\frac{\rho_{\rm end}}{M_P^4} \right)^{(1-\frac3k)} \right]^\frac12 \beta^{-\frac{3k-9}{k+2}} \, ,
\label{slimsc}
\eeq
and for $k=6$ becomes
\beq
\beta > 2.6 \times 10^{-4}  \sigma_{\rm eff}^{-8/9} \qquad {\rm or} \qquad \sigma_{\rm eff} > 9.3\times 10^{-5}  \beta^{-9/8}\, .
\label{blimsc6}
\eeq
Once again, the limit on the reheating temperature is given by Eq.~(\ref{trhbe}).

Some of these results are summarized in Tables \ref{table:1} and \ref{table:2}. In Table \ref{table:1}, 
we provide values of $\lambda$ from Eq.~(\ref{eq:normalizationlambda}) and $\phi_{\rm end}$ from Eq.~(\ref{phiend}) for $k = 4, 6, 8,$ and 10. Recall that $\rho_{\rm end} = V(\phi_{\rm end})$. The values of $\beta = a_\beta/\aend$ are taken from the numerical work described in the next section. Using these values we provide the lower limits on the effective couplings, $y_{\rm eff}$, $\mu_{\rm eff}$, and $\sigma_{\rm eff}$ as well as the lower limit on $\Trh$ to ensure that reheating occurs before the fragmentation of the condensate for $k = 4, 6, 8$ and 10. 
 
\begin{table}[!ht]
\centering
\begin{tabular}{||c c c c c c||} 
 \hline
 $k$ & $\lambda$  &$\phi_{\rm end}/M_P$ & $-\dot \phi_{\rm end}/(10^{-6}M_P^2)$ & $a_\beta/a_\text{end}$ & $a_\text{br}/a_\text{end}$ \rule[-1.ex]{0pt}{3.5ex}\\
 \hline\hline
 4 &  $3.42 \times 10^{-12}$ & 1.524 & 3.392 & 180 & 265  \\ 
 6 & $5.70 \times 10^{-13}$ & 1.983 & 3.332 & $4.5 \times 10^4$  & $3.3 \times 10^4$ \\
 8 & $9.51 \times 10^{-14}$ & 2.322 & 3.309 & $ 6 \times 10^6$ & $ 2.9 \times 10^6$\\
 10 & $1.58 \times 10^{-14}$ & 2.588 & 3.299 & $7 \times 10^8$ & $3.4 \times 10^8$\\
 \hline
\end{tabular}
\caption{Numerical values obtained for $\phi$ and $\dot \phi$ at the end of inflation for $V(\phi)$ given by Eq.~(\ref{eq:attractor}). They are input variables for lattice simulations
which give the fragmentation time $a_\beta$. $\lambda$ and $\phi_{\rm end}$ are determined from Eqs.~(\ref{eq:normalizationlambda}) and (\ref{phiend}) respectively. Also presented is the predicted backreaction time $a_{\rm br}$ discussed in Appendix \ref{Sec:simulation}.}
\label{table:1}
\end{table}

\begin{table}[!ht]
\centering
\begin{tabular}{||c c c c c||} 
 \hline
 $k$ & $y_{\rm eff}$  &$\mu_{\rm eff}$  &$\sigma_{\rm eff}$ &$T_{\rm RH}$\rule[-1.ex]{0pt}{3.5ex}\\
 \hline\hline
 4 & $1.61 \times 10^{-1}$  & $3.57 \times 10^{10}$ GeV   & $3.57 \times 10^{-6}$ & $1.14 \times 10^{13}$ GeV\\
 6 & $1.58 \times 10^{-2}$  & $1.84 \times 10^{5}$ GeV   & $5.37 \times 10^{-10}$ & $1.19 \times 10^{10}$ GeV\\
 8 & $1.32 \times 10^{-3}$ & $6.33 \times 10^{-1}$ GeV  & $ 9.59 \times 10^{-15}$ & $1.50 \times 10^{7}$ GeV \\
 10 & $3.62 \times 10^{-5}$ & $1.49 \times 10^{-6}$ GeV & $6.47 \times 10^{-20}$ & $1.80 \times 10^{4}$ GeV\\
 \hline
\end{tabular}
\caption{Lower limits on the effective couplings, $y_{\rm eff}$, $\mu_{\rm eff}$, and $\sigma_{\rm eff}$ to ensure that reheating occurs before the fragmentation of the condensate for $k = 4, 6, 8$ and 10. Also given is the corresponding lower limit on the reheating temperature so that reheating takes place before fragmentation.   }
\label{table:2}
\end{table}

Some comments on these limits are in order. Starting with the decay to fermions, we see that the lower limit to $y_{\rm eff}$ ranges from $\mathcal{O}(.1)$ to $\mathcal{O}(10^{-5})$ for $k = 4 - 10$.  However for effective couplings this large, we see from Eq.~(\ref{eq:yeffdef}) that the Lagrangian coupling $y$ must 
be very large (non-perturbative) since $\alpha_y \propto \mathcal{R}^{-1/2}$. In other words, for $y\lesssim 1$, we expect that
$y_{\rm eff} \lesssim 10^{-4}$ for $\mathcal{R} \gtrsim 10^{16}$
as seen in Fig.~\ref{fig:R}, which is the value of the kinetic 
factor at values of
$a_\beta/\aend\gtrsim 100$ when fragmentation becomes effective. Thus the fragmentation of the condensate is expected to always occur before reheating if the reheating is dominated by decays to fermion pairs.
In sharp contrast, the decay to boson pairs is not impeded, and the lower limit to $\mu$ is weaker than the limit to $\mu_{\rm eff}$ seen in Table \ref{table:2}. This is because $\mu_{\rm eff} \propto \mathcal{R}^{1/4} \mu$. Finally, although the effective coupling for scattering is also suppressed, $\sigma_{\rm eff} \propto \mathcal{R}^{-1/4} \sigma$, the limits on $\sigma_{\rm eff}$ are small enough that they may be satisfied for perturbative values of $\sigma$. 

In the case of scattering to scalars, if the limit on $\beta$ (or $\sigma$) is violated, reheating still 
occurs, but with an altered reheating temperature as the 
energy density of the 
inflaton quanta now scales as $a^{-4}$ rather than $a^{-6k/(k+2)}$. As a 
consequence the radiation energy density will scale as $a^{-3}$ as opposed to $a^{-18/(k+2)}$ when $ a_\beta < a < a_{\rm RH}$. 
If we ignore fragmentation, the reheating temperature produced from inflaton scatterings is
obtained from Eq.~(\ref{trh1}) using $\gamma_\phi$ from Eq.~(\ref{Eq:gammaphibis}) which corresponds to $\phi \phi \to b b$
\begin{align}
T_{\rm RH} \;&=\; \left(\frac{30}{g_\rho \pi^2} \right)^{\frac{1}{4}}
\left[\frac{\sqrt{3} k \sigma_{\rm eff}^2}{8 \pi (2k-5) (k(k-1))^\frac32 \lambda^\frac3k} 
\right]^{\frac{k}{4k-12}} M_P \, ,
\label{trhs}
\end{align}
which for $k=6$ simplifies to 
\beq
\Trh = \frac{1}{2 \sqrt{70} \pi} \left( \frac{3}{g_\rho \lambda} \right)^\frac14 \sigma_{\rm eff} M_P \simeq 0.0079 \,\frac{\sigma_{\rm eff}}{\lambda^\frac14} M_P \simeq 9.1\, \sigma_{\rm eff} M_P \, ,
\eeq
for $g_\rho=100$ and $\lambda = 5.7 \times 10^{-13}$.

It is relatively straightforward to determine the reheating temperature when $\sigma$ 
violates the condition in Eq.~(\ref{blimsc}). For $a < a_\beta$, the evolution of $\rho_\phi$ and $\rho_R$ is unaffected by fragmentation. For $a > a_\beta$, we can write
\begin{eqnarray}
\rho_\phi & = & \rho_{\rm end} \beta^{-\frac{6k}{k+2}}\left( \frac{a_\beta}{a} \right)^4 \\
\rho_R & = & \left[\frac{\sqrt{3} k \sigma_{\rm eff}^2}{8 \pi (2k-5) (k(k-1))^\frac32 \lambda^\frac3k} 
\right] \left( \frac{\rho_{\rm end}}{M_P^4}\right)^{\frac3k-1} \beta^{-\frac{18}{k+2}} \left( \frac{a_\beta}{a} \right)^3 \, .
\end{eqnarray}
Setting $\rho_\phi = \rho_R$, we can determine $a_\beta/\arh$ and $\rho_R(\arh)$. The correct reheating temperature, $\Trh^{(\beta)}$ in this case is related to that in Eq.~(\ref{trhs})
by
\beq
\Trh^{(\beta)} = \left[\frac{\sqrt{3} k \sigma_{\rm eff}^2}{8 \pi (2k-5) (k(k-1))^\frac32 \lambda^\frac3k} 
\right]^{1-\frac{k}{4k-12}} \beta^{\frac{9(k-4)}{2(k+2)}} \left( \frac{\rho_{\rm end}}{M_P^4}\right)^{\frac3k-\frac34} \Trh \, ,
\eeq
and for $k=6$,
\beq
\Trh^{(\beta)} = 0.019 \, \sigma_{\rm eff} \lambda^{-\frac14} \beta^\frac98 \left( \frac{\rho_{\rm end}}{M_P^4}\right)^{-\frac14} \Trh \simeq 1.1\times 10^4 \, \sigma_{\rm eff} \, \beta^\frac98 \, \Trh \, ,
\label{trhb}
\eeq

\begin{figure}[!ht]
\centering
    \includegraphics[width=0.65\textwidth]{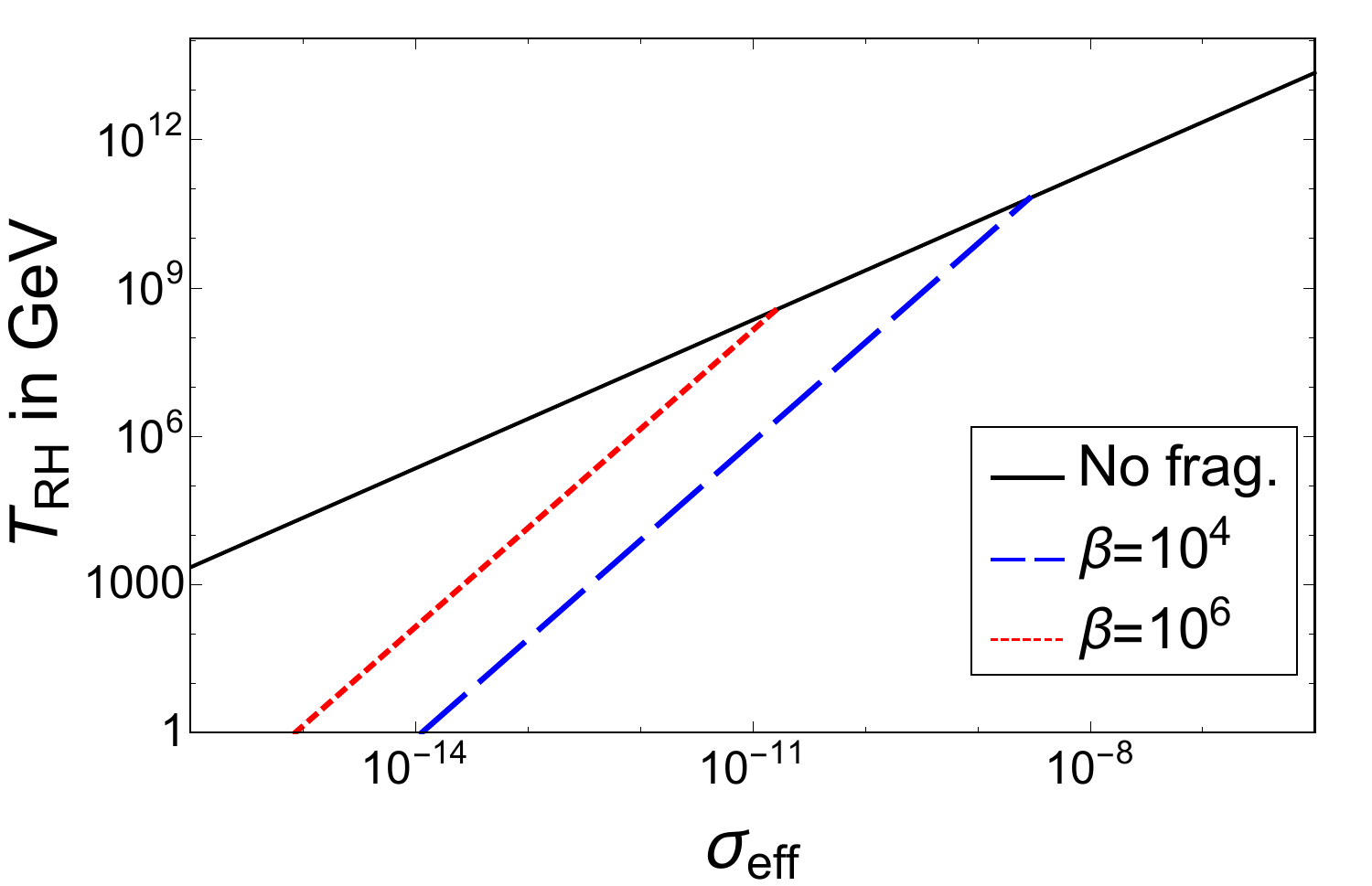}
    \caption{The reheating temperature $\Trh$ as a function of $\sigma_{\rm eff}$ for $k = 6$. The black curve shows the temperature with the effects of fragmentation ignored. The blue (dashed) and red (dotted) lines show the temperature $\Trh^{(\beta)}$ for $\beta = 10^4$  and $10^6$ respectively. }
    \label{scatplot}
\end{figure}

Figure \ref{scatplot} shows the effect of fragmentation on the reheating temperature for $k=6$. The black curve shows the reheating temperature as a function of the effective quartic coupling $\sigma_{\rm eff}$ using Eq.~(\ref{trhs}) which ignores the effects of fragmentation. The blue (dashed) and red (dotted) curves are computed using Eq.~(\ref{trhb}) for $\beta = 10^4$ and $10^6$ respectively. 
We clearly observe the change of slope when $\sigma_{\rm eff}$ drops to values below the limit given by Eq. (\ref{blimsc6}), passing
from $\Trh\propto \sigma_{\rm eff}$ to $\Trh\propto \sigma_{\rm eff}^{2}$.
The effect on the reheating temperature is more pronounced for smaller value of $\beta$, 
i.e. for earlier fragmentation.
On the other hand, for sufficiently
large $\sigma_{\rm eff}$ satisfying Eq.~(\ref{blimsc6}), the effects of fragmentation can be neglected.

\section{Fragmentation: a numerical approach}
\label{sec:frag}

\subsection{Generalities}

A non-quadratic scalar potential such as (\ref{eq:Vphi}) 
describes a self-interacting inflaton field. This self-interaction will generically 
transfer some of the energy of the coherent, oscillating 
inflaton field into inhomogeneous fluctuations $\delta\phi(t,\bx)$, through 
processes such as $\phi^{k-2}\rightarrow \delta \phi\, \delta \phi$. At linear 
order, the growth of these inhomogeneities can be 
described as the exponential excitation of the momentum 
modes of $\phi$ due to the presence of parametric 
resonances. The equation of 
motion that is satisfied by the inflaton perturbations is
\beq\label{eq:deltaphieom}
\ddot{\delta\phi} + 3H\dot{\delta\phi} - \frac{\nabla^2 \delta\phi}{a^2} + k(k-1)\lambda M_P^2\left(\frac{\phi(t)}{M_P}\right)^{k-2} \,\delta\phi \;=\; 0\,.
\eeq
The solution to this equation is not $\delta\phi=0$, as the fluctuations have a quantum mechanical origin, and therefore a non-vanishing initial vacuum value. Switching to conformal time $d\tau=dt/a$, and introducing the canonically normalized inflaton fluctuation $X=a\,\delta\phi$, we can write
\beq\label{eq:Xdef}
X(\tau,\bx) = \int \frac{\diff  ^3\bp}{(2\pi)^{3/2}}\,e^{-i\bp\cdot\bx} \left[ X_p(\tau)\hat{a}_{\bp} + X_p^*(\tau)\hat{a}^{\dagger}_{-\bp} \right]\,.
\eeq
Here we have introduced the comoving momentum $\bp$, and a set of annihilation and creation operators $\hat{a}_{\bp}$ and $\hat{a}^{\dagger}_{\bp}$ which satisfy the canonical commutation relations $[\hat{a}_{\bp},\hat{a}^{\dagger}_{\bp'} ] = \delta(  \bp-\bp')$, $[\hat{a}_{\bp},\hat{a}_{\bp'} ] = [\hat{a}^{\dagger}_{\bp},\hat{a}^{\dagger}_{\bp'} ] = 0$. The Wronskian constraint

\beq
X_pX^{*\prime}_p - X_p^*X_p' \;=\; i
\eeq
(where ${}^{\prime}$ denotes differentiation with respect to conformal time) is imposed to ensure that the canonical commutation relations between $X_p$, and its momentum conjugate, $X_p'$, are also fulfilled. 

Substitution of (\ref{eq:Xdef}) into (\ref{eq:deltaphieom}) yields the following equation for the dynamics of the individual mode functions, 
\beq\label{eq:eomX}
X_{p}'' + \omega_p^2 X_{p} \;=\; 0\,,
\eeq
and the effective frequency is given by
\beq\label{eq:omega}
\omega_p^2 \;\equiv\; p^2 - \frac{a''}{a} + k(k-1)\lambda a^2 M_P^2\left(\frac{\phi(t)}{M_P}\right)^{k-2}\,.
\eeq
The initial condition for the modes is given by the Bunch-Davies vacuum solution,
\beq
X_p(\tau_0) \;=\; \frac{1}{\sqrt{2\omega_p}} \,, \quad X_p'(\tau_0) \;=\; -\frac{i\omega_p}{\sqrt{2\omega_p}}\,,
\eeq
imposed at the beginning of reheating for all sub-horizon modes. In the absence of expansion, Eq.~(\ref{eq:eomX}) contains a periodic effective frequency, and is a special case of Hill's equation~\cite{Magnus2004-br}. The Floquet theorem guarantees that the solution has in general the following form,
\beq
X_{p}(\tau) \;=\; e^{\mu_p \tau}g_1(\tau) + e^{-\mu_p \tau}g_2(\tau)\,,
\eeq
where $g_1(\tau)$ and $g_2(\tau)$ are periodic functions, and $\mu_p$ are known as the Floquet 
exponents. For a given value of the model couplings, these 
exponents will have non-vanishing real parts for 
certain values of the comoving momentum $\bp$ 
(resonance bands). Therefore, these solutions will grow 
exponentially fast, manifesting the phenomenon of 
parametric resonance. The important point here is that even for weak couplings $\lambda$, 
this resonance can be strong enough to eventually bring 
the system into the non-linear regime.\footnote{For 
more details on the structure of the resonance bands for 
monomial potentials, see~\cite{Lozanov:2017hjm}.} Quantitative details can be found in Appendix~\ref{Sec:simulation}. Alternatively, 
this resonance is manifested as the exponential growth of 
the occupation numbers of the modes, defined as
\beq
n_{p} \;=\; \frac{1}{2\omega_p}\left| \omega_p X_p - i X'_p \right|^2\,.
\eeq

When we account for expansion, the oscillating 
part of the frequency (third term on the right-hand side 
of (\ref{eq:omega})) will redshift as $a^{-4\frac{k-4}{k+2}}$. Thus, conversion of the homogeneous condensate 
into an inhomogeneous gas of free inflatons is expected to occur on a larger time scale for larger values of $k>4$. Nevertheless, 
as we will now show, for all the values of $k>4$ that we 
consider, the non-linear regime is reached. The 
backreaction of the inhomogeneous component on 
the inflaton condensate ``fragments'' it, partially 
replacing the classical oscillating inflaton by free 
inflaton quanta with non-vanishing momentum. It is 
worth noting that for $k=4$, the strength of the resonance 
does not decrease with time. The exponential growth of 
fluctuations is only moderated by the non-linear 
mode-mode interactions, and the condensate component 
continues to be depleted after the onset of 
backreaction~\cite{Garcia:2023eol}.

\subsection{Lattice simulations} 

Using \CL~\cite{Figueroa:2020rrl, Figueroa:2021yhd}, we performed 
lattice simulations to obtain $\beta$ for different values 
of $k$ in Eq.~(\ref{eq:attractor}). 
Prior 
to simulations, we evolved the system numerically until the 
end of inflation corresponding to the 
condition $\dot \phi_\text{end}=-\sqrt{V(\phi_\text{end})}$. Numerical evaluations of the 
inflaton field and its time derivative at the end of 
inflation are provided in Table~\ref{table:1}. This yields an energy density at the end of inflation $\rho_{\rm end} \sim 5.6 - 5.8 \times 10^{62}~\textrm{GeV}^4$ mildly dependent on $k$.

We describe the details
of our methodology used to perform the lattice simulations of the fragmentation
in Appendix \ref{Sec:simulation}.
On the lattice, the full non-linear partial differential equation for the space-time dependent inflaton field (\ref{eq:deltaphieom})
is then solved over a configuration-space lattice. The energy density of the inflaton is computed from the spatial average of the energy-momentum tensor of $\phi$, which we denote with an over-bar,
\beq
\rho_{\phi} \;=\; \overbar{\frac{1}{2}\dot{\phi}^2 + \frac{1}{2a^2}(\nabla\phi)^2+V(\phi)}\,.
\label{Eq:spatialaverage}
\eeq
 For definiteness
 we define the condensate component of the energy density as follows,
\beq
\rho_{\bar \phi} \;=\; \frac{1}{2}\dot{\bar{\phi}}^{\,2} + V(\overbar{\phi})\,,
\eeq
that is, the energy density of the {\it spatially averaged inflaton field} which 
is then different from the spatially average energy density of the inflaton field (\ref{Eq:spatialaverage}).

\begin{figure}[!ht]
\includegraphics[width=\textwidth]{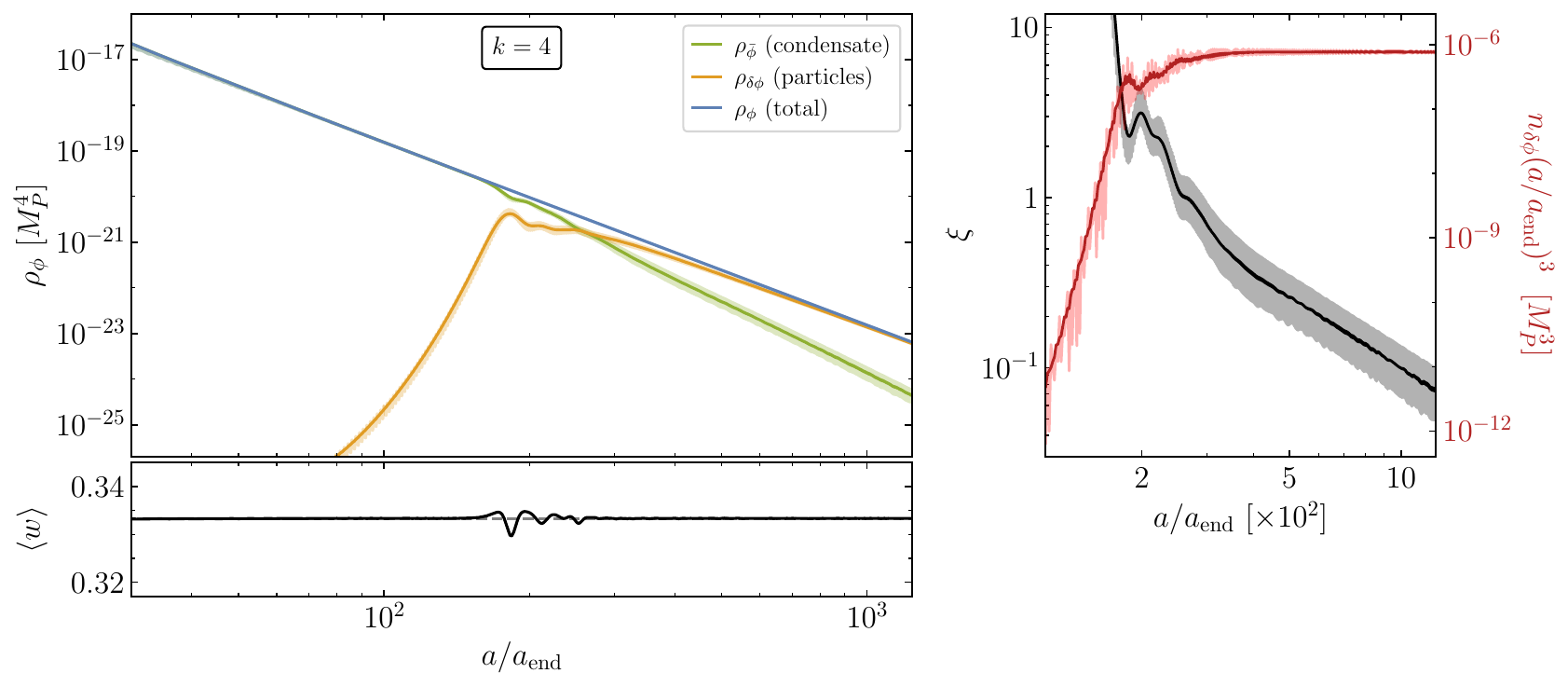}\par\noindent
\includegraphics[width=\textwidth]{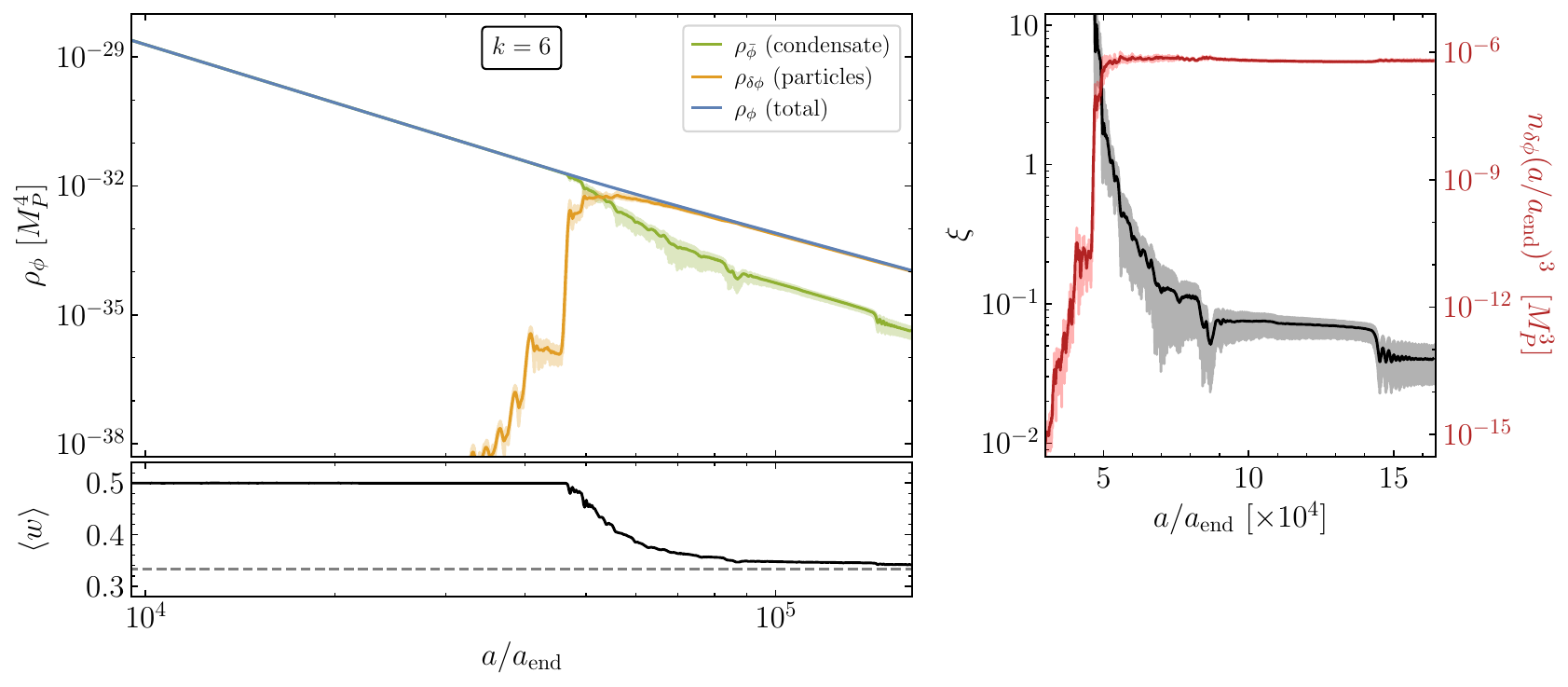}
\caption{Left:~inflaton energy density in the classical condensate (${\rho_{\bar \phi}}$, green), in free particles ($\rho_{\delta\phi}$, yellow) and the sum of both ($\rho_{\phi}$, blue), as functions of the scale factor. Right:~ratio of energy densities $\xi={\rho_{\bar \phi}}/\rho_{\delta\phi}$ (black) and comoving number density of inflatons $n_{\delta\phi}(a/a_{\rm end})^3$ (red), during and after fragmentation. Bottom:~the oscillation-averaged equation of state.}\label{energydensityplot1}
\end{figure}

\begin{figure}[!ht]
\includegraphics[width=\textwidth]{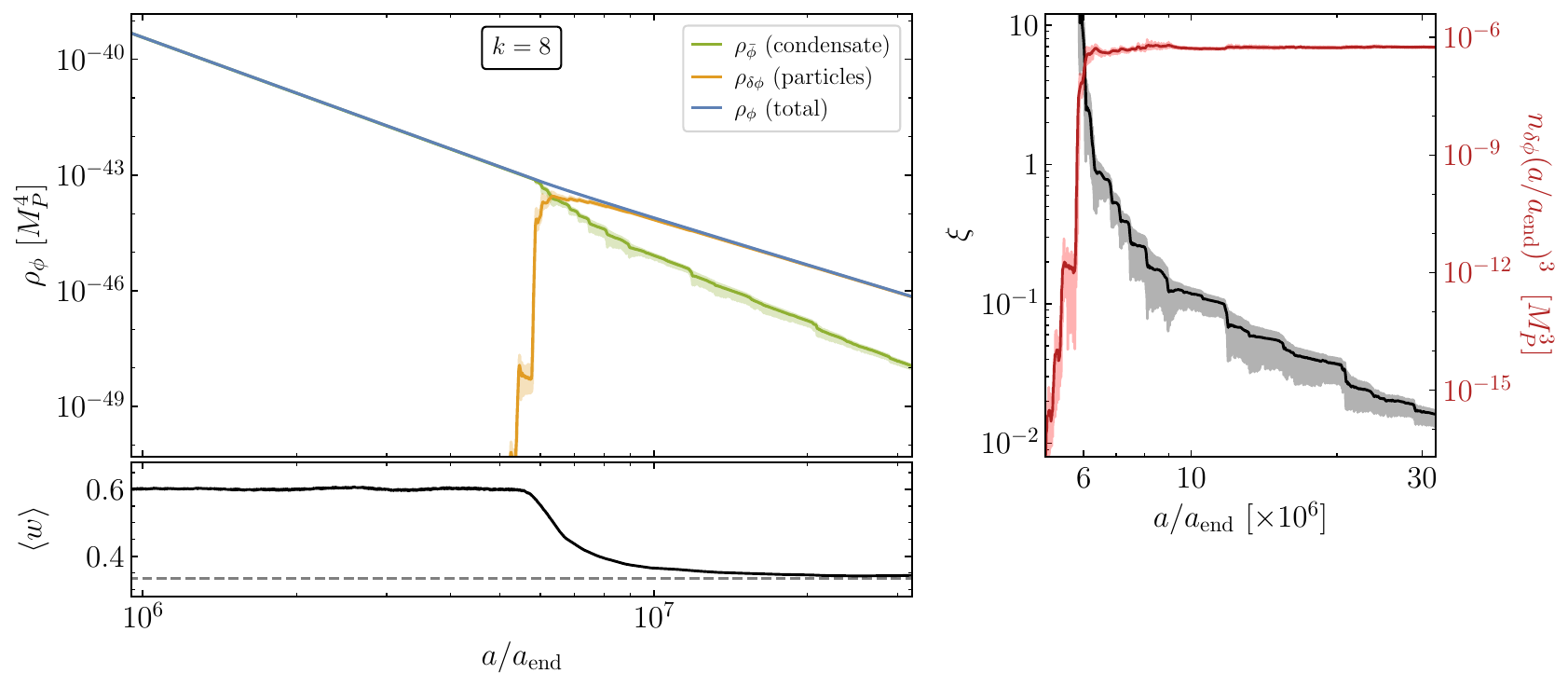}\par\noindent
\includegraphics[width=\textwidth]{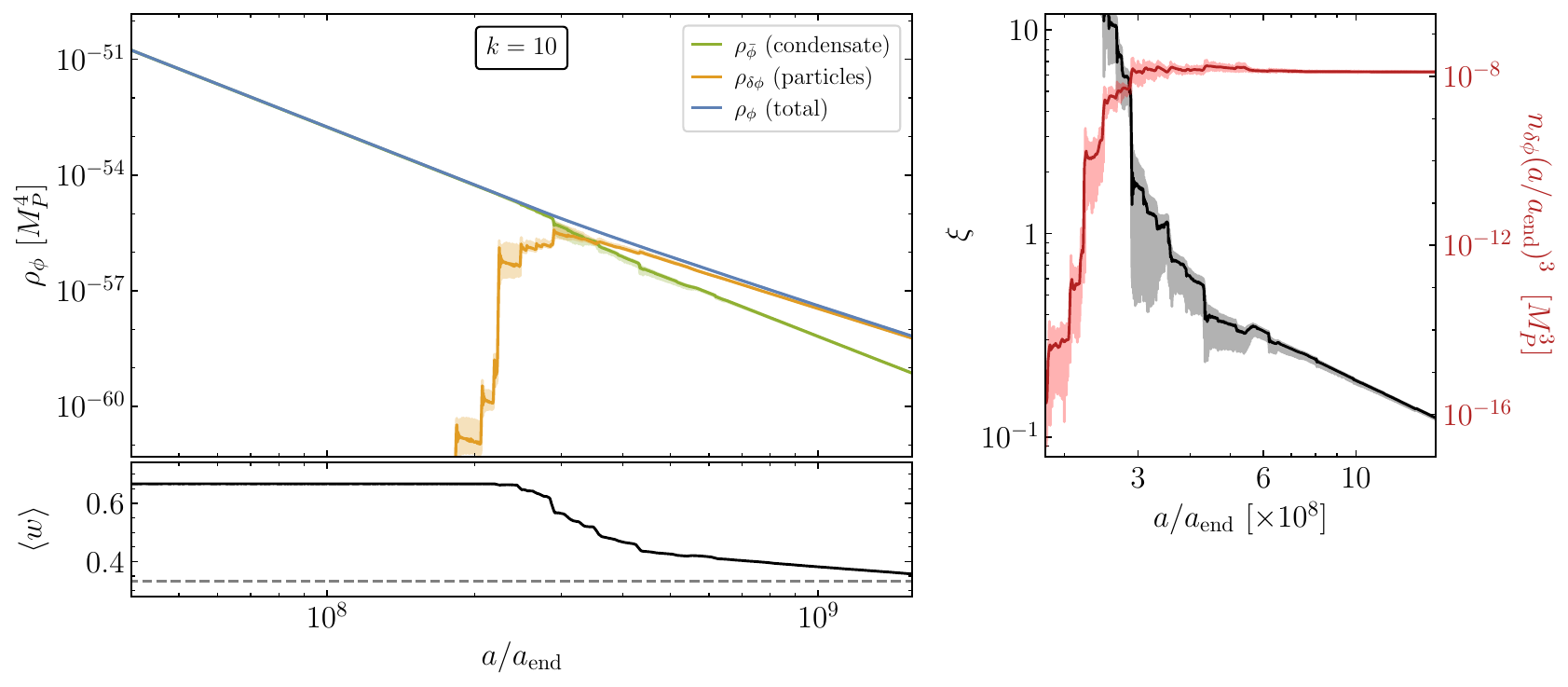}
\caption{Left:~inflaton energy density in the classical condensate (${\rho_{\bar \phi}}$, green), in free particles ($\rho_{\delta\phi}$, yellow) and the sum of both ($\rho_{\phi}$, blue), as functions of the scale factor. Right:~ratio of energy densities $\xi={\rho_{\bar \phi}}/\rho_{\delta\phi}$ (black) and comoving number density of inflatons $n_{\delta\phi}(a/a_{\rm end})^3$ (red), during and after fragmentation. Bottom:~the oscillation-averaged equation of state.}\label{energydensityplot2}
\end{figure}

 Spectral data in turn is obtained upon Fourier transformation of configuration-space quantities. 
Two important quantities can be obtained from the 
phase space distribution (PSD), the number density and the energy density of 
the fluctuations. Their UV-regular forms are computed as
\begin{align}\label{eq:ndeltaphi}
n_{\delta\phi} \;&=\; \frac{1}{(2\pi)^3a^3}\int \diff^3\bp\, n_p\,,\\ \label{eq:rhodeltaphi}
\rho_{\delta\phi} \;&=\; \frac{1}{(2\pi)^3a^4}\int \diff^3\bp\, \omega_p n_p\,,
\end{align}
the second of which corresponds to $\rho_{\delta\phi}=\rho_{\phi}-\rho_{\bar \phi}$~\footnote{ It was shown in Ref.~\cite{Garcia:2023eol}, corresponding to the case $k=4$, that estimating $\rho_{\delta\phi}$ from Eq.~(\ref{eq:rhodeltaphi}) numerically is in excellent agreement with estimating $\rho_{\delta\phi}=\rho_{\phi}-\rho_{\bar \phi}$ directly from the lattice.}.
These quantities are represented in Figs.~\ref{energydensityplot1} and \ref{energydensityplot2}
for $k=4$, 6, 8 and 10,
in addition to the averaged 
equation of state parameter. 
Notice that the process of self-fragmentation does not 
occur instantaneously. As observed in 
Figs.~\ref{energydensityplot1} and \ref{energydensityplot2}, most of the energy density of $\phi$ is in the condensate initially, and 
consequently the mean equation of state parameter 
follows the relation (\ref{eq:wk}). After the 
exponential growth of inhomogeneities, $\langle w\rangle$ has a transient 
period of irregular behavior (for $k=4$), or starts 
decreasing as the particle 
component of $\rho_{\phi}$ 
starts to dominate. 
Then the convergence toward an equation of state corresponding
to a gas of relativistic inflaton ($w=1/3$) is relatively fast.
Considering the potential 
sources of discrepancies, 
such as different initial 
conditions 
and numerical uncertainties, 
our results appear to be 
consistent with those of Ref.~\cite{Lozanov:2017hjm}.
The appendix provides details of
 an analytical approximation of the typical resonantly enhanced scale, Eq. (\ref{krange}), which is in good agreement with our estimate. 
For any value of $k$,
only $\sim 2$ e-folds after the beginning of the fragmentation, a large majority of the condensate has already been transformed into 
particles.
We also show the time evolution of the inflaton fragmented quanta 
occupation number 
in the right panels of Figs.~\ref{energydensityplot1} and \ref{energydensityplot2} as well as $\xi \equiv \rho_{\bar{\phi}} /\rho_{\delta \phi}$, the ratio of the density in the remaining condensate to that in particles from the curves in the left panels. 

\subsection{Effect of the fragmentation on $\Trh$}

The analytic limits derived in Section \ref{sec:analim}, were based on the assumption that when fragmentation of the condensate occurs, it is complete
and all of the energy density in the condensate is converted to free and massless (for $k \ge 4$) particles which as a result do not decay. This would preclude the possibility of reheating. 
While this assumption is roughly 99\% correct, as seen in Figs.~\ref{energydensityplot1} and \ref{energydensityplot2}
where $\rho_{\bar{\phi}} \gtrsim 0.01 \times\rho_{\delta \phi}$ at the end of the fragmentation process,
the 1\% difference  from our assumption in Section \ref{sec:analim} can significantly affect our conclusions.

Indeed, in Figs.~\ref{energydensityplot1} and \ref{energydensityplot2}, we see not only the onset of fragmentation, but also the `late' time evolution of both the condensate and the free inflaton-particles produced from the destruction of the condensate.
For each of the four examples shown $k=4, 6, 8$ and 10
we note that 
the ratio of the energy density remaining in the condensate 
relative to the energy density in particles, $\xi$, decreases slowly after fragmentation as seen in the right panels of Figs.~\ref{energydensityplot1} and \ref{energydensityplot2}. 
Thus while the amplitude 
of the inflaton oscillations associated with the condensate is 
drastically reduced when 
fragmentation occurs, it is not driven to {\it exactly} 0. 
The presence of a relic condensate 
generates a mass term for the $\phi$-particles which allows them to decay.
From Eq.~(\ref{Eq:meanrhophi}), we can still write 
${\rho_{\bar\phi}} = \lambda M_P^4\left(\frac{\phi_0}{M_P}\right)^k$ and 
we obtain for the amplitude
\beq
\frac{\phi_0}{M_P} = \left( \frac{{\rho_{\bar \phi}}}{\lambda M_P^4} \right)^{\frac{1}{k}}\,,
\eeq
and therefore the interaction between the $\phi$-particles and the condensate provides an effective mass term 
\beq\label{eq:mcondf}
m_\phi = \sqrt{k(k-1)} \lambda^{1/2} \left(\frac{\phi_0}{M_P} \right)^{\frac{k-2}{2}} M_P = \sqrt{k(k-1)} \lambda^{1/k} \left(\frac{{\rho_{\bar \phi}}}{M_P^4} \right)^{\frac{k-2}{2k}} M_P
\propto a^{-3\frac{k-2}{k+2}}\,.
\eeq
This mass term may then allow the inflaton to decay (and thus  reheat) if its width dominates
the expansion rate, $\Gamma_\phi \gtrsim H$.
After fragmentation occurs, the Hubble rate, dominated by the relativistic fragmentation products,
redshifts as $H \propto a^{-2}$.
The
$\phi$-particles of number density $n$ have mean
energy $\overline E=\frac{\rho_{\rm tot}}{n} \simeq (\pi^\frac72/30^\frac34 \zeta(3)) \rho_{\delta\phi}^\frac14 \equiv c_e \rho_{\delta\phi}^\frac14$ with $c_e \approx 3.57$.  After fragmentation, these particles are relativistic and their decay rate is suppressed by a time dilation factor $\simeq m_\phi/\overline E$.\footnote{A more rigorous treatment would consist in estimating the boost factor for a given mode from the occupation number $n_p$, which is beyond the scope of this paper. For a full analysis based on the Boltzmann equation for $k=4$ and fermionic decays, see~\cite{Garcia:2023eol}. More intuitively, this factor corresponds to the Lorentz boost that one needs to account for in order to express in the cosmic comoving frame, the decay rate computed in the mean rest frame of inflaton fluctuations, which is of order $\sim m_\phi/\bar E$. }

For the case of inflaton decay to fermions, 
as we will show, the width 
$\Gamma_\phi \sim \frac{m_\phi^2}{\bar E}\propto a^{-\frac{5k-14}{k+2}}$. We then see that for $k\ge 6$,
the width redshifts as fast or faster than the Hubble 
rate which is $\propto a^{-2}$, and the relativistic particles 
never decay: reheating is never completed. Their 
lifetime increases faster than the age of the Universe.
In this case, the analytical results 
we obtained in Section~\ref{Sec:decaytofermions} are still valid.
Especially Eq.~(\ref{Eq:minimaltrhfermions}) which requires a minimal reheating temperature so that reheating is complete {\it before} fragmentation occurs.
However as previously argued, this requires a non-perturbative coupling unless the effects of kinematic blocking can be avoided. 

To be more precise,  we can compute the expected reheating temperature making use of the numerical results discussed above which provide the ratio of the energy density in the residual condensate to the total energy density dominated by the inflaton fluctuation, $\delta \phi$. 
For fermionic final states, the rate for particle decay is given by\footnote{Note that in this case, the reheating is due to the decay of the dominant component which is the bath of $\phi$-particles whose coupling with the fermions is $y$ and not $y_{\rm eff}$.} Eq.(\ref{Eq:gammaphi}) multiplied by the time dilation factor
\beq
\Gamma_\phi = \frac{y^2 m_\phi^2}{8 \pi \overline E}  = c_k \xi^{\frac{k-2}{k}} \beta^{\frac{k-4}{2(k+2)}} \left( \frac{\rho_\text{end}}{M_P^4} \right)^{\frac{3k-8}{4k}} \left( \frac{a_\text{end}}{a} \right)^{\frac{5k-14}{k+2}} M_P \, ,
\label{gamfragf}
\eeq
with $c_k \equiv \frac{y^2}{8\pi c_e} k(k-1) \lambda^{\frac{2}{k}}$.
To arrive at Eq.~(\ref{gamfragf}), we used 
${\rho_{\bar \phi}} = \xi \rho_\text{end} (a_\text{end}/a)^{6k/(k+2)}$ and $\overline E = c_e \rho_{\delta\phi}^\frac14 = c_e \rho_\text{end}^\frac14 \beta^{\frac{(4-k)}{2(k+2)}}(a_\text{end}/a)$.
If we define the value of the scale factor at reheating from $\Gamma_\phi = H$, we find
\beq
\frac{\arh}{\aend}= (\sqrt{3} c_k)^{\frac{k+2}{3(k-6)}} \left(\frac{\rho_\text{end}}{M_P^4}\right)^{\frac{(k+2)(k-8)}{12k(k-6)}} \xi^{\frac{k^2-4}{3k(k-6)}}\beta^{\frac{k-4}{2(k-6)}}  \, ,
\eeq
where we used $H=\rho_{\delta \phi}^\frac12/\sqrt{3}M_P$.
The energy density at $a_{\rm RH}$ is then
\bea
&&
\rho_{\rm RH}= \rho_\text{end} \beta^{\frac{8-2k}{k+2}} \left( \frac{\aend}{\arh} \right)^4 = M_P^4 (\sqrt{3} c_k \xi^{\frac{k-2}{k}})^\frac{k+2}{3(6-k)} \left(\frac{\rho_\text{end}}{M_P^4}\right)^{\frac{2k^2-12k+16}{3k(k-6)}}
\beta^{-4\frac{(k-2)(k-4)}{(k+2)(k-6)}}
\nonumber
\\
&&
\Rightarrow 
\Trh^{(\beta)}= \left(\frac{30}{g_\rho \pi^2} \right)^{\frac{1}{4}} M_P(\sqrt{3} c_k \xi^{\frac{k-2}{k}})^\frac{k+2}{12(6-k)} \left(\frac{\rho_\text{end}}{M_P^4}\right)^{\frac{2k^2-12k+16}{12k(k-6)}}
\beta^{-\frac{(k-2)(k-4)}{(k+2)(k-6)}}
\label{Eq:trhfermions}
\eea
As one can see from Eq.~(\ref{gamfragf}), the decay rate redshifts as $a^{-\frac{5k-14}{k+2}}$ which for $k=6$ is $a^{-2}$ and thus redshifts in parallel with the Hubble rate and reheating can not occur.  
Actually for $k=6$, it is necessary to redo the integration in Eq.~(\ref{eq:aeom}) which produces a log contribution so that $\Gamma_\phi \sim a^{-2} \ln^\frac12 a$,  i.e.~barely faster than the $\delta\phi$ redshift. 

\begin{figure}[!ht]
\centering    \includegraphics[width=0.95\textwidth]{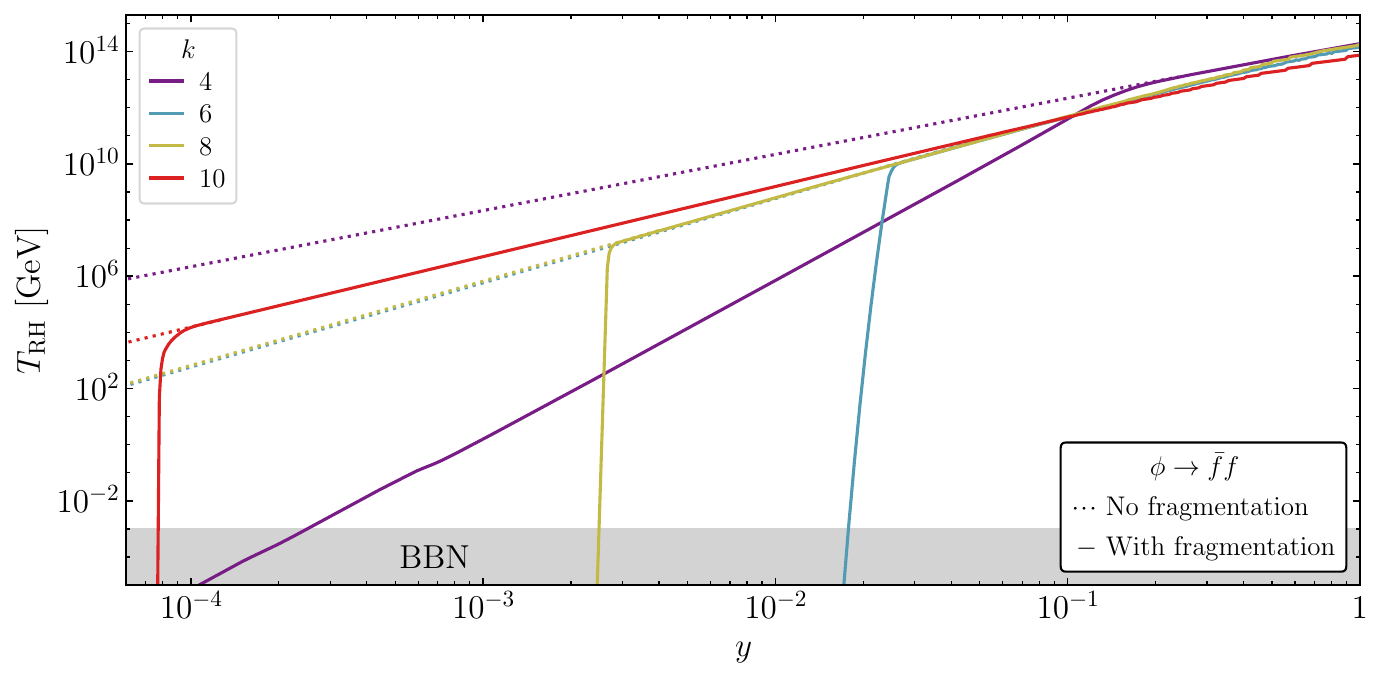}
    \caption{The reheating temperature as a function of the inflaton coupling to fermions, $y$. The dotted lines show the dependence when fragmentation is ignored. These are given by either Eq.~(\ref{trh1}) for $k= 4, 6$ or Eq.~(\ref{trh1-}) for $k= 8, 10$. The solid curves include the effects of fragmentation. When $y$ (or $y_{\rm eff})$ satisfies the limits in Table \ref{table:2} the solid and dotted curves track each other. At lower $y$, reheating no longer occurs. The effects of kinematic blocking are not included here.}
    \label{TrehF}
\end{figure}

 To illustrate our point, we show in Fig.~\ref{TrehF} the reheating temperature as a function of the coupling $y$ for four values of $k$, where we have neglected the kinematic blocking factor $\mathcal{R}$.
The dotted lines show the dependence of $\Trh$ when fragmentation is ignored. These are given by either Eq.~(\ref{trh1}) for $k= 4,6$ with $\Trh \propto y_{\rm eff}^\frac{k}{2}$ or Eq.~(\ref{trh1-}) for $k= 8, 10$ with $\Trh \propto y_{\rm eff}^\frac{3k}{2k-8}$.
 Since the (log)slopes for $k = 6$ and 8 are equal, the two lines are indistinguishable on the scale of the figure (they are not, however identical). The limits from Table \ref{table:2} are easily read from the figure once the relation between
 $y$ and $y_{\rm eff}$ from Fig.~\ref{alphaplot} is used. For example, for $k=6$,
above $y \simeq 0.2$, we have the result from section \ref{Sec:decaytofermions} (the limit seen here corresponds to the limit $y_{\rm eff} > 0.016$ found in Table \ref{table:2}).
At lower values of $y$ there is a very rapid decline in the reheating temperature. Due to the aforementioned log correction to the dependence on the scale factor, the curve is not quite vertical. 
For larger $k$, any departure from vertical below the limit on $y$ is numerical. 
When the effects of fragmentation are included, we see from 
Eq.~(\ref{Eq:trhfermions})
that $\Trh^{(\beta)}\propto y^{\frac{k+2}{6(6-k)}}$.
For $k > 6$, the decay rate (\ref{gamfragf}) would redshift faster than the Hubble rate ($a^{-2}$) and the limits from section \ref{Sec:decaytofermions} must be obeyed\footnote{The leftover condensate could still decay, but will not be able to inject an energy to the radiation bath comparable to $\rho_{\delta\phi}$.}  and even then perturbative couplings are only possible if the effects of kinematic blocking can be avoided.\footnote{One possibility is an  inflaton-fermion axial coupling of the kind $\mathcal{L} \, \supset \, \phi \bar f \gamma_5 f$ }
The case of $k=4$ is special, since its conformal nature 
leads to a continuous 
conversion of the condensate into inhomogeneities even 
past the onset of 
fragmentation. Nevertheless, the decay into fermions after 
backreaction is not as 
strongly suppressed, and 
allows for reheating, as 
discussed in detail in~\cite{Garcia:2023eol}.

For the decay to scalars, the 
rate is enhanced due to fragmentation as it is 
proportional to $m_\phi^{-1}$ which is decreased. This 
remains true even when the 
effect of time dilation is taken into account.  But the 
rate is no longer enhanced by the effective mass, $\mu_{\rm eff}^2$ and as one can see 
from Eq.~(\ref{Eq:gammaphi}) where $l<0$ 
for $k>2$ the decay rate is now
given by
\beq
\Gamma_\phi = \frac{\mu^2}{8 \pi \overline E} = \frac{\mu^2}{8 \pi c_e M_P^2} \left(\frac{\rho_\text{end}}{M_P^4} \right)^{-\frac14} \beta^{\frac{k-4}{2(k+2)}} \left( \frac{a}{a_\text{end}} \right) M_P \, .
\eeq
Note that there is no longer a dependence on $\xi$. 
In this case, we have
\beq
\frac{\arh}{\aend}=\left( \frac{\sqrt{3} \mu^2}{8 \pi c_e M_P^2} \right)^{-\frac13} \left(\frac{\rho_\text{end}}{M_P^4} \right)^{\frac14} \beta^{-\frac{k-4}{2(k+2)}}
\eeq
implying a very simple result for $\rho_{\rm RH}$ which is independent of $k$ and $\beta$ as well
\bea
&&
\rho_{\rm RH}= M_P^4 \left( \frac{\sqrt{3} \mu^2}{8 \pi c_e M_P^2} \right)^{\frac43}
\nonumber
\\
&&
\Rightarrow 
\Trh^{(\beta)}= \left(\frac{30}{g_\rho \pi^2} \right)^{\frac{1}{4}} M_P   \left(\sqrt{3} \frac{\mu^2}{8 \pi c_e M_P^2} \right)^{\frac13}
\label{trhsb}
\eea

\begin{figure}[!ht]
\centering
\includegraphics[width=0.95\textwidth]{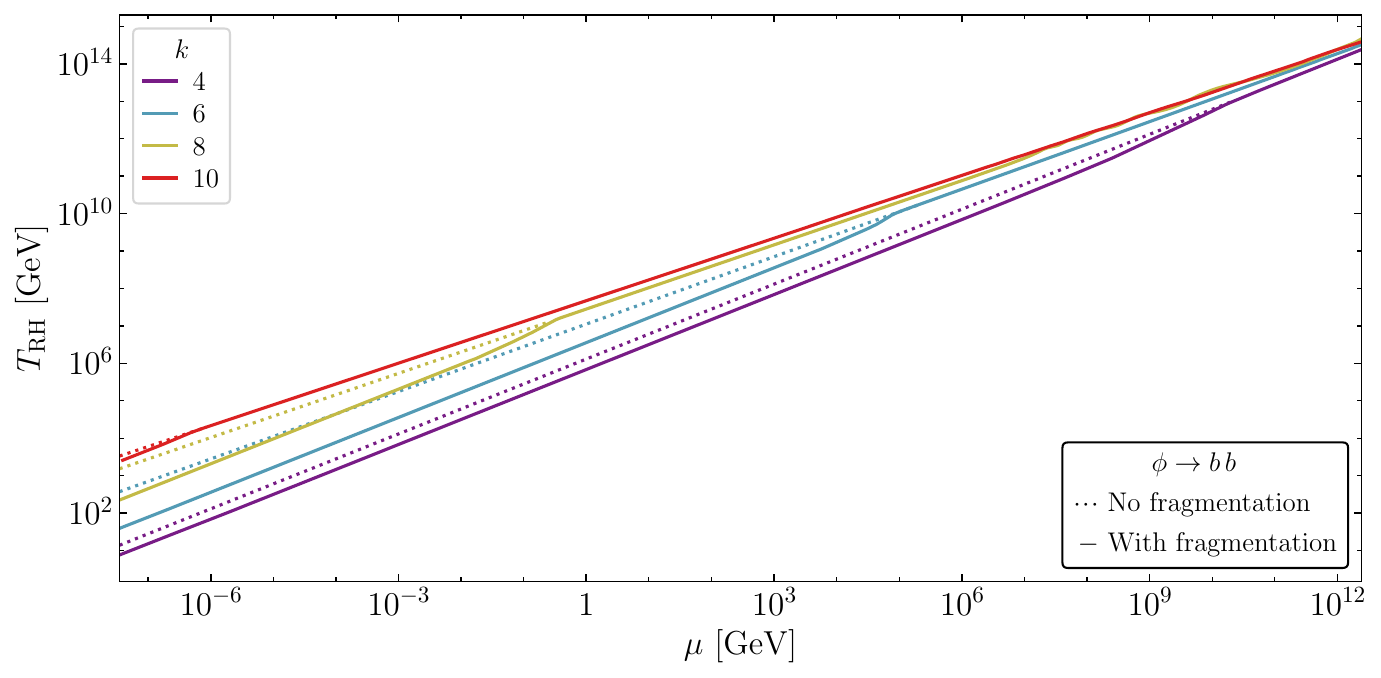}
    \caption{The reheating temperature as a function of the inflaton coupling to bosons, $\mu$. The dotted lines show the dependence when fragmentation is ignored. These are given by Eq.~(\ref{trh1}). The solid curves include the effects of fragmentation. When $\mu$ (or $\mu_{\rm eff})$ satisfies the limits in Table \ref{table:2} the solid and dotted curves track each other. At lower $\mu$, the reheating temperature is suppressed given in Eq.~(\ref{trhsb}). The effects of kinematic blocking are not included here.}
    \label{TrehS}
\end{figure}

In this case, for $k=6$, the ratio of the reheating temperature including 
fragmentation to that neglecting fragmentation is 
\beq
\frac{\Trh^{(\beta)}}{\Trh} = \left(\frac{\sqrt{3}}{8 \pi c_e}\right)^\frac13 \left(\frac{13}{6} 8 \pi \sqrt{10} \right)^\frac{3}{10} \lambda^\frac{1}{20} \left(\frac{\mu}{M_P}\right)^\frac23 \left(\frac{M_P}{\mu_{\rm eff}}\right)^\frac35 \simeq 0.3 \left(\frac{\mu}{M_P}\right)^\frac23 \left(\frac{M_P}{\mu_{\rm eff}}\right)^\frac35
\eeq
which is slightly less than 1 and decreases with decreasing $\mu$.
This can be seen in Fig.~\ref{TrehS} 
which as in Fig.~\ref{TrehF} shows the dependence of $\Trh$ on $\mu$ in the absence of fragmentation (dotted lines) and with the effects of fragmentation included (solid lines). In this case kinematic enhancements are also neglected. 
From Eq.~(\ref{trh1}), we see that $\Trh \sim \mu^\frac{k}{2(k-1)}$ for the dotted lines, and from Eq.~(\ref{trhsb}), $\Trh^{(\beta)} \sim \mu^\frac23$. For $k = 4$, the two 
slopes are the same, but at the 
values of $\mu$ in Table \ref{table:2}, there is a 
suppression at lower $\mu$ since $\mu < \mu_{\rm eff}$. For larger $k$, the slopes of the dotted lines 
decrease where the slopes of the 
solid lines are constant. This is 
apparent in the figure. Clearly the 
effects of fragmentation are mild compared with those found for decays to fermions. 

\begin{figure}[!ht]
\centering
\includegraphics[width=0.95\textwidth]{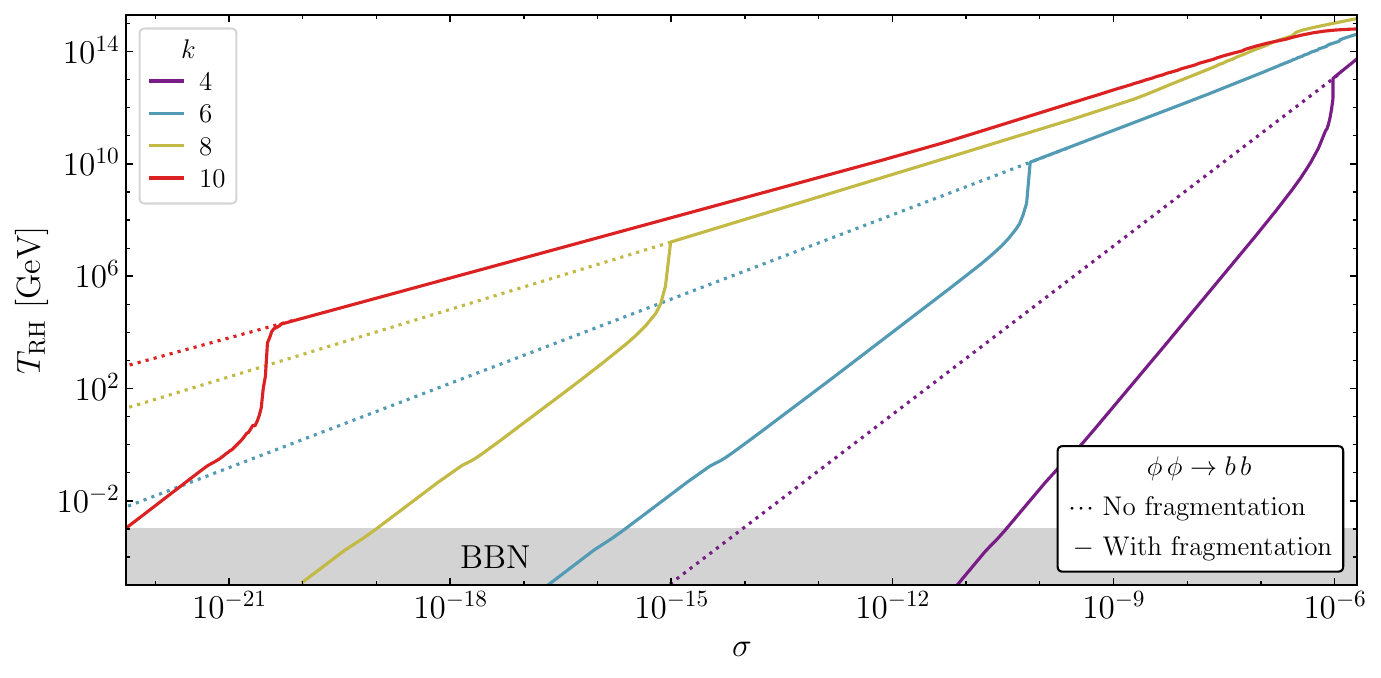}
    \caption{ {The reheating temperature as a function of the inflaton coupling to bosons, $\sigma$. The dotted lines show the dependence when fragmentation is ignored. These are given by Eq.~(\ref{trh1}). The solid curves include the effects of fragmentation. When $\sigma$ (or $\sigma_{\rm eff})$ satisfies the limits in Table \ref{table:2} the solid and dotted curves track each other.   At lower $\sigma$, the reheating temperature is suppressed given in Eq.~(\ref{eq:TRHbetasigma}). The effects of kinematic blocking are not included here.}}
    \label{Trehsigma}
\end{figure}

Finally, in the case of inflaton scatterings to bosons, 
we can write
the scattering rates as
\beq
\Gamma_\phi=n\frac{\sigma^2}{8 \pi \overline E^2}=\frac{\tilde c \sigma^2}{8 \pi}\rho_{\rm tot}^\frac{1}{4}
\eeq
with $\tilde c=\frac{900\times 30^{\frac{1}{4}}\zeta(3)^3}{\pi^{10}\sqrt{\pi}} \approx 0.022$. The condition 
$\Gamma_\phi(\arh)=H(\arh)$ then gives
\beq
\frac{\arh}{\aend}=\frac{8 \pi}{\sqrt{3}\tilde c \sigma^2}
\left(\frac{\rho_\text{end}}{M_P^4}\right)^\frac{1}{4}\beta^{\frac{4-k}{2(k+2)}}\,,
\eeq
and
\bea
&&
\rho_{\rm RH}=M_P^4\left(\frac{\sqrt{3} \sigma^2 \tilde c}{8\pi}\right)^4
\\
&&
\Rightarrow \Trh^{(\beta)}=M_P\left(\frac{30}{g_\rho \pi^2}\right)^\frac{1}{4}\left(\frac{\sigma^2 \tilde c \sqrt{3}}{8 \pi}\right)\,,
\label{eq:TRHbetasigma}
\eea
which is again independent of $\xi, \beta$, and $k$. 
For $k=6$, the ratio of the reheating temperature including 
fragmentation to that neglecting fragmentation is
\beq
\frac{\Trh^{(\beta)}}{\Trh}=\frac{\sigma^2}{\sigma_{\rm eff}}
\frac{(10 \lambda)^\frac{1}{4} \sqrt{210}}{4\sqrt{\pi}} \tilde c\,.
\eeq
This is illustrated in Fig.~\ref{Trehsigma} 
which as in Fig.~\ref{TrehF} and Fig.~\ref{TrehS} shows the dependence of $\Trh$ on $\sigma$ in the absence of fragmentation (dotted lines) and with the effects of fragmentation included (solid lines). In this case kinematic suppression is also neglected. At large values of $\sigma$ the solid lines keep track of the dashed lines corresponding to $\Trh \sim \mu^\frac{6k}{4k-16}$ from Eq.~(\ref{trh1-}) (for $k>4$), i.e. the slopes tend to decrease with $k$.  At small values of $\sigma$, from Eq.~(\ref{eq:TRHbetasigma}) $\Trh^{(\beta)} \sim \sigma^2$ independently of $k$ (for $k>4$) giving rise to identical slopes for the solid lines corresponding to $k=6,8,10$ as
apparent in the figure. At small values of $\sigma$, the error made on the reheating temperature by neglecting fragmentation can reach several orders of magnitude. Clearly the 
effects of fragmentation in this case are stronger than the effect on decays to scalars, but milder compared to decays into fermions.

As a conclusion, we see that the effect of fragmentation on the reheating process is {\it highly} dependent on the decaying mode. While the fermionic
channel $\phi \rightarrow \bar ff$ is almost excluded due to 
the conjunction of inefficient decay relative to the expansion rate and strong 
kinematic suppression, the decay $\phi \rightarrow bb$ is almost unaffected, whereas the 
scattering process 
$\phi \phi \rightarrow b b$ leads to a 
significantly lower reheating temperature.

\section{Summary}
\label{sec:disc}

In most simple, single field inflation models, the energy density of the universe is dominated by the inflaton scalar potential $V(\phi)$. As inflation ends, this nearly constant energy density is transferred in part to the  kinetic energy associated with an oscillating inflaton condensate. To recover a standard hot big bang universe, 
the inflaton must to some degree couple to Standard Model fields so that the process of reheating occurs. 
In the simplest case, where the inflaton potential is quadratic near its minimum, inflaton decays can reheat the Universe to a temperature $\Trh \propto y$, where $y$ is the coupling to SM fields. However even in this simple case, when examined in detail, the reheating process is more involved.
Early decay products can thermalize and achieve a temperature $T_{\rm max}$ far greater than $\Trh$, though their total contribution to the energy density may be small. As inflaton decays continue, in the case of a quadratic potential the energy density of the newly formed radiation bath scales as $\rho_R \sim a^{-3/2}$ while the energy density of the inflaton condensate redshifts as $\rho_\phi \sim a^{-3}$. This continues until ``reheating" when $\rho_R = \rho_\phi$.
Soon thereafter $\rho_\phi$ drops off exponentially as decays become rapid and $\rho_R \sim a^{-4}$ typical of a radiation dominated universe. 

This picture is altered when
the shape of the potential about the minimum is no longer quadratic. When the potential can be expressed as $V \sim \phi^k$ near its minimum, the energy density of the condensate scales as $\rho_\phi \sim a^{-6k/(k+2)}$
and the behavior of the energy of the radiation scales as either Eq.~(\ref{Eq:rhoRapp})
or (\ref{Eq:rhoRappter}) and depends on $k$, and whether the decay products are bosons or fermions (or if two-body scatterings dominate the reheating process). These differences affect the maximum temperature and the reheating temperature. 

For $k \ge 4$, in addition to producing SM fields, scatterings can produce inflaton fluctuations which behave as a gas of massless inflaton particles. The case for $k=4$ was recently treated in \cite{Garcia:2023eol}. For $k>4$, it was argued~\cite{Copeland:1995fq,Salmi:2012ta,Amin:2010dc,Amin:2011hj,Lozanov:2016hid,Lozanov:2017hjm} that the equation of state parameter $w = (k-2)/(k+2)$ would evolve from $w>1/3$ to $w=1/3$ as the condensate fragments through the appearance of solitonic objects called oscillons. The complete destruction of the condensate to a gas of massless inflatons would preclude reheating if reheating does not complete before the fragmentation process begins. 

In this work, we have first derived analytical limits to the inflaton-SM couplings such that reheating occurs before the fragmentation of the condensate assuming that fragmentation occurs when $a = \beta \aend$. For decays to SM fermions, we found limits on the effective couplings which are relatively large.
So large, in fact, that the corresponding Lagrangian couplings were necessarily non-perturbative unless the effects of kinematic blocking could be evaded. In contrast, for decays to bosons or scatterings to bosons, limits to either the effective (dimensionful) coupling (in the case of decays) or quartic coupling (in the case of scatterings), were relatively easy to satisfy, particularly for larger values of $k$. We noted that in the case of scatterings, even if the limit on the coupling is violated (and fragmentation occurs before reheating), reheating still occurs through the scattering of now massless inflaton particles as shown in Fig.~\ref{scatplot}.

For a more quantitative approach, we performed a numerical study of the fragmentation process by simulating the post-inflation dynamics with the public code \CL. From the simulations, for each value of $k$ we kept track of the energy density stored in the form of the inflaton homogeneous (condensate) and inhomogeneous (inflaton quanta) components as illustrated in Figs.~\ref{energydensityplot1} and \ref{energydensityplot2}. We determined the instant of fragmentation $\beta$ and found that fragmentation occurs later for larger values of $k$:  $a_\beta/a_\text{end}\simeq \{180,4.5 \times 10^4,  6 \times 10^6,7\times 10^8\}$ for $k=\{4,6,8,10\}$.
More importantly, from the simulations we found that the fragmentation process is not completed. 
Although typically $\sim 99\% $ of the energy density is transferred to inflaton particles, $\mathcal{O}(1\%)$ of the total energy density remains in the form of a homogeneous condensate. This remaining component is important as it induces an effective mass to the fragmented inflaton particles potentially allowing them to decay even after the bulk of the fragmentation process is complete. 

We find that the effects of fragmentation on the reheating process depend strongly on the decaying mode. In the case of 
decays to fermions, for $k>6$
($w>1/2$), the inflaton decay rate redshifts faster than the expansion rate and post-fragmentation decays do not occur (as originally argued in Section~\ref{Sec:decaytofermions}). For $k=6$, decays drop off dramatically when the limit in Eq.~(\ref{bylimf}) is violated as seen in Fig.~\ref{TrehF}. For decays to scalars via a dimensionful coupling, decays do indeed continue subsequent to fragmentation and the resulting reheating temperature is independent of $k$ and the specific value of $\beta$. The resulting reheating temperature is only slightly suppressed when fragmentation effects are included as seen in Fig.~\ref{TrehS}. For scattering to scalars via a quartic coupling, fragmentation does not prevent reheating but substantially diminishes the efficiency of the process. For small couplings, the resulting reheating temperature can be reduced by several orders of magnitude as illustrated in Fig.~\ref{Trehsigma}.

\section*{Acknowledgments}
The authors would like to thank Daniel Figueroa, Simon Cl\'ery, and Essodjolo Kpatcha for very fruitful discussions. This project has received funding /support from the European Union's Horizon 2020 research and innovation program under the Marie Sklodowska-Curie grant agreement No 860881-HIDDeN. This work was performed using HPC resources from the “M\'esocentre” computing center of CentraleSup\'elec and \'Ecole Normale Supérieure Paris-Saclay supported by CNRS and R\'egion \^Ile-de-France (http://mesocentre.centralesupelec.fr/). The work of K.A.O. was supported in part by DOE grant DE-SC0011842 at the University of Minnesota. M.A.G.G. is supported by the DGAPA-PAPIIT grant IA103123 at UNAM, and the CONAHCYT ``Ciencia de Frontera'' grant CF-2023-I-17. M.P. acknowledges support by the Deutsche Forschungsgemeinschaft (DFG, German Research Foundation) under Germany’s Excellence Strategy – EXC 2121 “Quantum Universe” – 390833306. J.Y. would like to acknowledge helpful discussions with Mustafa A. Amin and his group members at Rice University.  The authors acknowledge the support of the Institut Pascal at Université Paris-Saclay during the Paris-Saclay Astroparticle Symposium 2023, with the support of the P2IO Laboratory of Excellence (program ``Investissements d'avenir'' ANR-11-IDEX-0003-01 Paris-Saclay and ANR-10-LABX-0038), the P2I axis of the Graduate School of Physics of Université Paris-Saclay, as well as IJCLab, CEA, IAS, OSUPS, and the IN2P3 master project UCMN. 

\appendix

\section*{APPENDIX}

\section{Simulating the fragmentation}
\label{Sec:simulation}

Our simulations are performed using 
\CL~\cite{Figueroa:2020rrl, Figueroa:2021yhd}.
We use the $\alpha-$time defined via $\diff \tau_\alpha =a^{-\alpha} \omega_* \diff t $ with
\begin{equation}
    \alpha\,=\,\frac{3(k-2)}{(k+2)}\, \qquad \qquad \qquad \omega_* = \sqrt{\lambda} M_P \,,
\end{equation}
and we define the normalized field $\tilde \phi = \phi / f_*$ and inflaton potential $V \equiv V/(f^2_* \omega^2_*) $ with $f_* = M_P$. By using the average EOS $w\simeq (k-2)/(k+2)$ prior to fragmentation and the expression of $\lambda$ in Eq.~(\ref{eq:normalizationlambda}), the $\alpha-$time can be related to the scale factor via 
\begin{equation}
   \tau_\alpha \, \simeq \, \frac{(k+2) \omega_* }{18 \pi  M_P } \, \left( \frac{A_s \tanh ^k\left(\frac{1}{4} \ln \left(\frac{1}{3} k \left(\sqrt{k^2+3}+k\right)+\frac{1}{2}\right)\right)}{N_*^2} \right)^{-1/2} \left[ \left( \dfrac{a}{a_\text{end}}  \right)^{\frac{6}{k+2}}-1 \right] \,,
\end{equation}
normalized to $\tau_\alpha(a_\text{end})=0$.

\begin{figure}
%\url{m1l0_xi50_ls1e-8_nk.pdf}\\
\includegraphics[width=0.43 \linewidth]{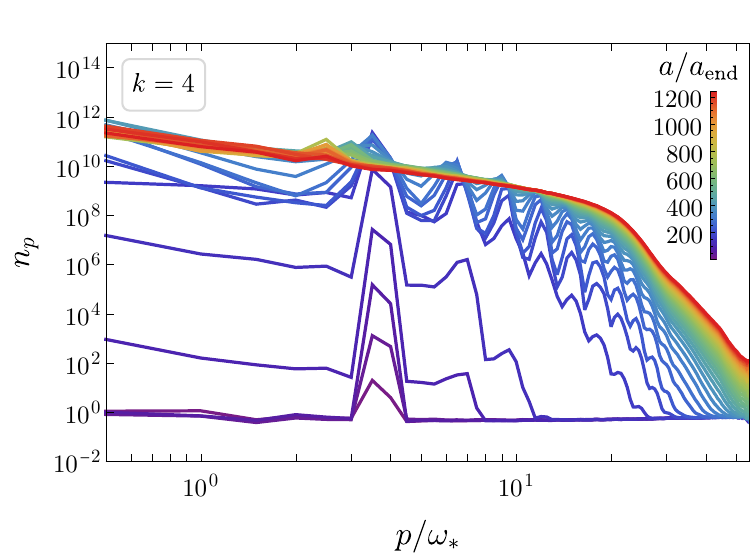}
\includegraphics[width=0.43 \linewidth]{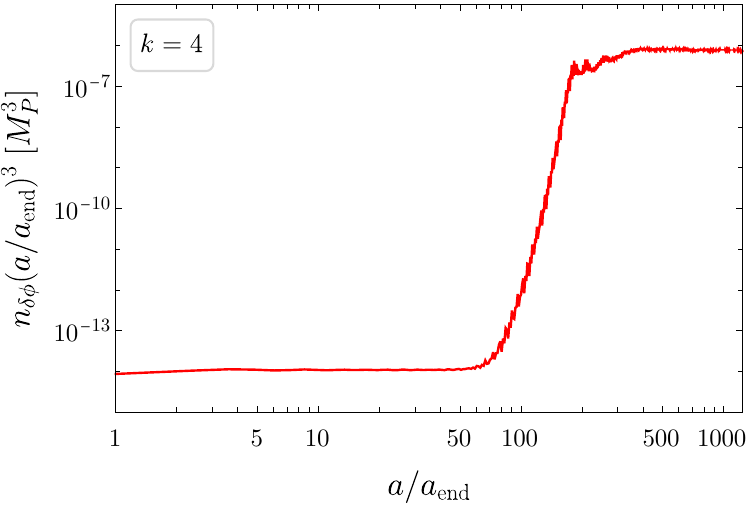}\\
\includegraphics[width=0.43 \linewidth]{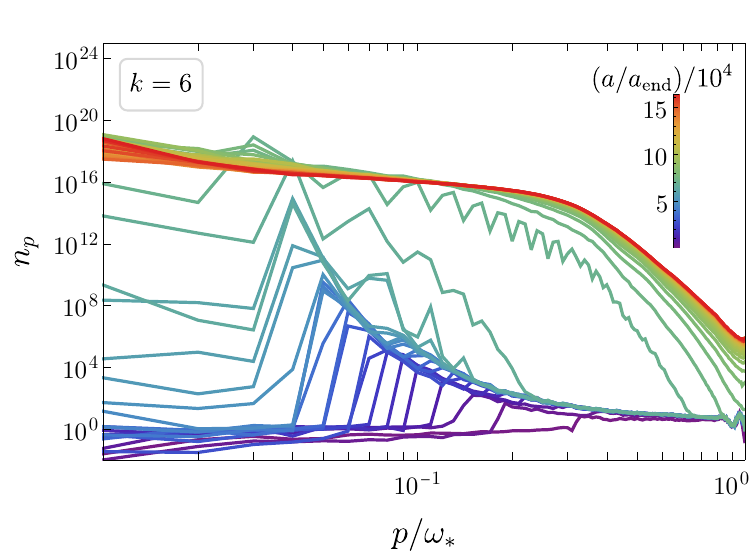}
\includegraphics[width=0.43 \linewidth]{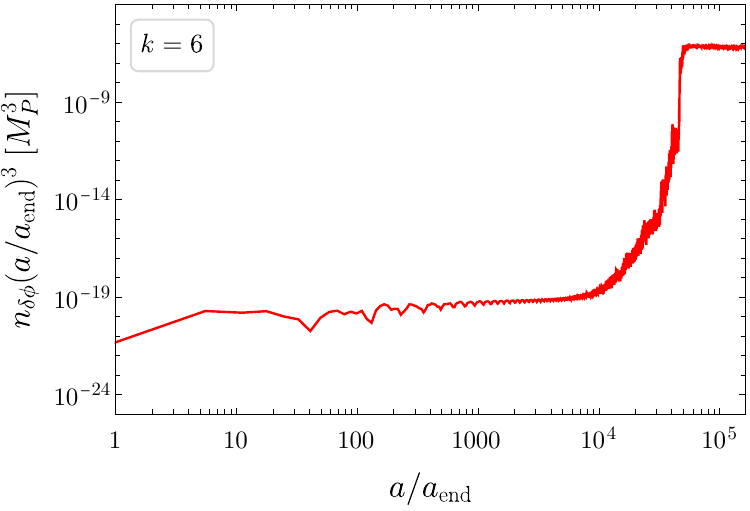}\\
\includegraphics[width=0.43 \linewidth]{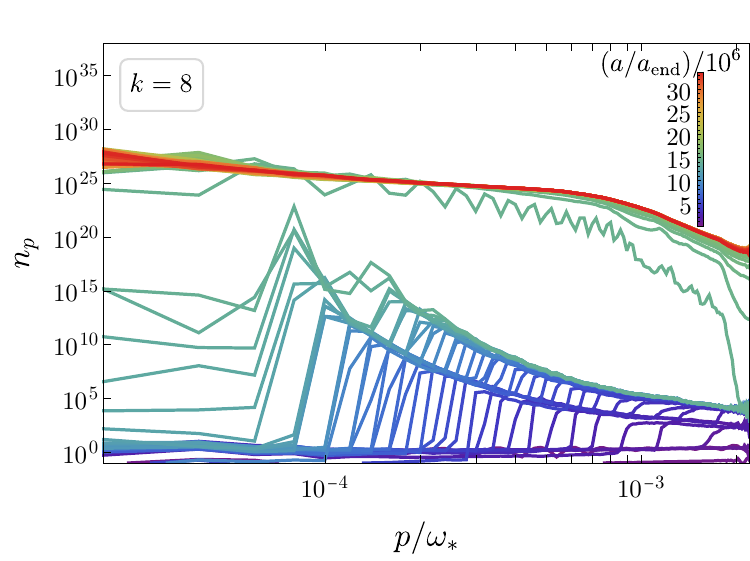}
\includegraphics[width=0.43 \linewidth]{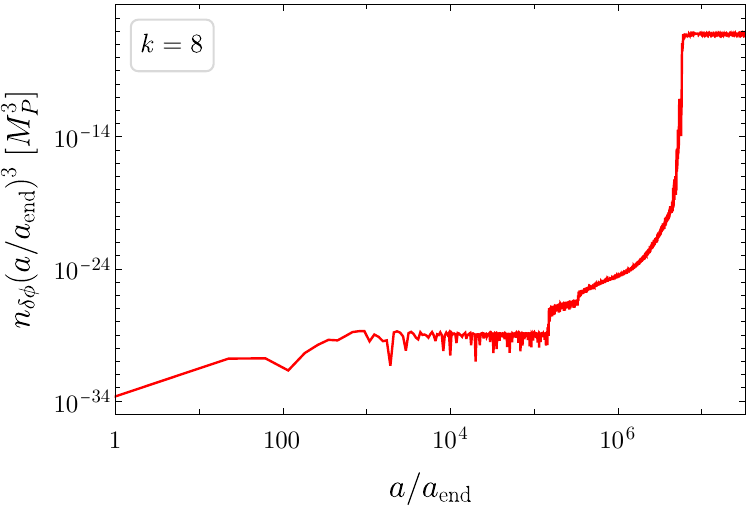}\\
\includegraphics[width=0.43 \linewidth]{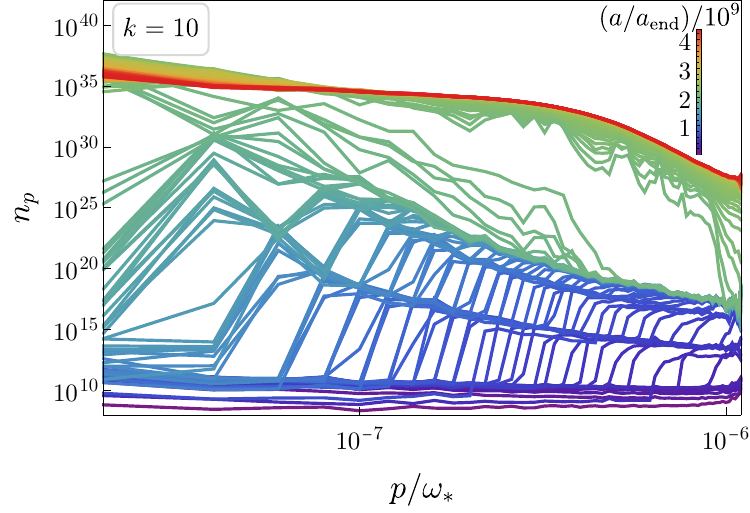}
\includegraphics[width=0.43 \linewidth]{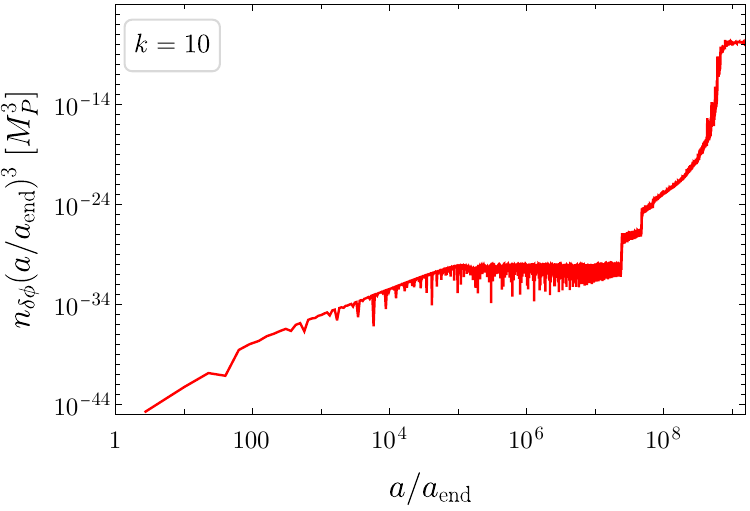}
\caption{Occupation number and comoving number density for the inflaton fluctuations. Both quantities are defined in Sec.~\ref{sec:frag}.}\label{numberdensityplot}
\vspace{-0.5cm}

\end{figure}

 According to \cite{Lozanov:2017hjm}, in the case of weak self-interaction, self-fragmentation is completed by the inflaton quanta produced on the first 
narrow instability band of the Floquet theorem. Consequently, the backreaction 
time, which corresponds to the moment 
when the generated particles start to affect the inflaton background, is 
expressed by the Floquet quotient and the width of the first narrow 
instability band. We represent the 
predicted backreaction time as $\beta^{\textrm{pred}}_{\textrm{br}} \equiv \frac{a_\textrm{br}}{a_\textrm{end}}$ and calculate it numerically using $\Delta N_{\rm{br}}^{\rm{pred}}\equiv\ln\left(\frac{a_{\text{br}}}{a_{\text{end}}}\right)\approx\frac{n+1}{3}\ln\left[\frac{1}{d\delta}\frac{\kappa}{\Delta\kappa}\frac{M}{m_{Pl}}\frac{|4-2n|}{n+1}\right]$ for $k \neq 4$ and $\ln\left(\frac{a_{\text{br}}}{a_{\text{end}}}\right)\approx\ln\left[\frac{1}{d\delta}\frac{M}{m_{Pl}}\right]$ for $k = 4$. $n$ corresponds to $k/2$ and $M$ to $\sqrt{6 \alpha} M_P$ with $\alpha=1$, etc. in our notation; see Fig. 4 and eq. (11) in \cite{Lozanov:2017hjm} for details.
The predicted back reaction time for several values of $k$ are listed in the last column of Table \ref{table:1}.

It is crucial for running lattice simulations to identify when and at what energy scale the physics event we want to observe occurs.
While important modes are 
produced at a constant $\kappa\equiv \bp/am$, in simulations, it is the comoving momentum $\bp$ that is fixed. 
In other words, observing $m^2 \propto k \lambda \phi^{k-2}$, $\tilde \phi \propto a^\frac{-6}{k+2}$ and $\kappa \sqrt{k}=\mathcal{O}(1)$ for the 
production events at the first narrow 
band allows us to predict the range of comoving momentum that we need to scan. Therefore, at backreaction time, we should focus on the following ranges of physics scale
\begin{eqnarray}
\bp/\sqrt{\lambda} M_P&\propto \tilde \phi_\textrm{end} \left(\frac{a_\textrm{br}}{a_\textrm{end}}\right)^{-3\frac{k-2}{k+2}+1}=\begin{cases}
1.524 \times 265^{-0} \sim \mathcal{O}(1) &\textrm{for}~ k=4\\
1.983 \times (3.3 \times 10^4)^{-0.5} \sim 1.1 \times 10^{-2} &\textrm{for}~ k=6\\
2.322 \times (2.9 \times 10^6)^{-0.8} \sim 1.6 \times 10^{-5} &\textrm{for}~ k=8\\
2.588 \times (3.4 \times 10^8)^{-1} \sim 7.6 \times 10^{-9} &\textrm{for}~ k=10 ~,
\end{cases}
\label{krange}
\end{eqnarray}
where we find that the typical momentum scale of the fragmentation is inversely proportional to $k$.

\begin{table}[!ht]
\centering
\begin{tabular}{||c c  c c||} 
 \hline
 $k$ & $p_\text{IR}/\omega_*$   & $N$ & $\Delta \tau_\alpha$ \rule[-1.ex]{0pt}{3.5ex}\\
 \hline\hline
 4 &  $0.5$ & $128$  & $10^{-3}$  \\ 
 6 & $10^{-2}$ &  $128$  & $10^{-3}$  \\
 8 & $2 \times 10^{-5}$  & $128$ & $10^{-3}$ \\
 10 & $2 \times 10^{-8}$  & $64$ & $10^{-3}$ \\
 \hline
\end{tabular}
\caption{Numerical input values used for the lattice simulations. $N$ is the number of lattice sites per dimension, i.e. corresponding to $N^3$ points on a cubic lattice. $\Delta \tau_\alpha$ is the ($\alpha-$time) time step. $p_\text{IR}/\omega_*$ is the lowest infrared Fourier scale that can be resolved on the lattice.}
\label{table:3}
\end{table}

As the resonance bands vary with time for $k>4$, one has to choose lattice parameters accordingly to fully capture the time-dependent resonance structure. For our simulations, we used the parameters given in Table~\ref{table:3}. The inflaton-fluctuation occupation number is shown in Fig.~\ref{numberdensityplot} (left panels) as a function of the Fourier scale for $k=4,6,8,10$ from top to bottom. The occupation numbers evaluated at different times are represented with different colors. One can clearly see from these plots that the resonance structure shifts towards the infrared for $k>4$ while the main resonance for $k=4$ remains at the same scale due to the (quasi) conformal invariance of the inflaton potential close to the minimum. One can also see from these plots that the lattice parameters given in Table.~\ref{table:3} allow for the resonance structure to be well captured for all cases. In the right panels of Fig.~\ref{numberdensityplot}, we represented the comoving number density of inflaton fluctuations, i.e. the occupation number integrated over momenta. One can see that once fragmentation is reached, the comoving number density freezes to a constant value in all cases considered.

%\addcontentsline{toc}{section}{References}
%\bibliographystyle{utphys}
%\bibliography{references} 

% %\bibliographystyle{utphys}
 \bibliographystyle{apsrev4-1}

% %\bibliography{biblio} 

\end{document}